\documentclass[sigconf]{acmart}

\usepackage{hyperref}[colorlinks,linkcolor=blue]
\usepackage{pdfpages}
\usepackage{ulem}
\usepackage[section]{placeins}
\usepackage{appendix}
\usepackage{multirow}
\usepackage{multicol}
\usepackage{graphicx}
\usepackage[format=hang,font=small,labelfont=bf]{caption}
\usepackage{booktabs}
\usepackage{caption} 
\usepackage{svg}
\usepackage{paralist}
\usepackage{wrapfig}
\usepackage{nameref}
\usepackage{makecell}
\AtBeginDocument{%
  }

\setcopyright{acmcopyright}
\copyrightyear{2018}
\acmYear{2018}
\acmDOI{XXXXXXX.XXXXXXX}

\acmConference[HHME 2023]{The $19^{th}$ Joint Academic Conference on Harmonious Human-Machine Environment}{August 25--27,
  2023}{Harbin, China}
\acmPrice{15.00}
\acmISBN{978-1-4503-XXXX-X/18/06}

\begin{document}

\title{ALens: An Adaptive Domain-Oriented Abstract Writing Training Tool for Novice Researchers}

\author{Chen Cheng}
\authornote{Both authors contributed equally to this research.}
\email{chengchen@shanghaitech.edu.cn}
\author{Ziang Li}
\authornotemark[1]
\email{liza1@shanghaitech.edu.cn}
\affiliation{%
  \institution{School of Information Science and Technology, ShanghaiTech University}
  \streetaddress{393 Middle Huaxia Road, Pudong}
  \city{Shanghai}
  \country{China}
  \postcode{201210}
}

\author{Zhenhui Peng}
\email{pengzhh29@mail.sysu.edu.cn}
\affiliation{%
  \institution{School of Artificial Intelligence, Sun Yat-Sen University}
  \streetaddress{Hanlin Road, Xiangzhou District}
  \city{Zhuhai}
  \country{China}}

\author{Quan Li}
\authornote{The corresponding author.}
\email{liquan@shanghaitech.edu.cn}
\affiliation{%
  \institution{School of Information Science and Technology, ShanghaiTech University Shanghai Engineering Research Center of Intelligent Vision and Imaging}
  \streetaddress{393 Middle Huaxia Road, Pudong}
  \city{Shanghai}
  \country{China}
  \postcode{201210}
}


\newcommand{\changed}{\textcolor{black}}
\newcommand{\ziang}{\textcolor{purple}}
\newcommand{\cc}{\textcolor{orange}}
\newcommand{\delete}{\textcolor{red}}
\newcommand{\finished}{\textcolor{gray}}
\begin{abstract}
The significance of novice researchers acquiring proficiency in writing abstracts has been extensively documented in the field of higher education, where they often encounter challenges in this process. Traditionally, students have been advised to enroll in writing training courses as a means to develop their abstract writing skills. Nevertheless, this approach frequently falls short in providing students with personalized and adaptable feedback on their abstract writing. To address this gap, we initially conducted a formative study to ascertain the user requirements for an abstract writing training tool. Subsequently, we proposed a domain-specific abstract writing training tool called \textit{ALens}, which employs rhetorical structure parsing to identify key concepts, evaluates abstract drafts based on linguistic features, and employs visualization techniques to analyze the writing patterns of exemplary abstracts. A comparative user study involving an alternative abstract writing training tool has been conducted to demonstrate the efficacy of our approach.
\end{abstract}



\maketitle

\section{Introduction}
\par Paper writing is an essential skill that junior graduate students or researchers should master~\cite{zhu2004faculty} because of its importance in learning, understanding, applying, and synthesizing new knowledge~\cite{dryer2013scaling}. Basically, the structure of a typical research paper follows a pattern known as ``\textit{King Model}''~\cite{derntl2014basics}, which delineates the thematic progression of an article through six sections: \textit{title}, \textit{abstract}, \textit{introduction}, \textit{body}, \textit{discussion}, and \textit{references}~\cite{derntl2014basics}. Among the major components of academic papers, abstracts, which usually consist of separate paragraphs outlining the content of the paper ~\cite{tullu2019writing}, have become increasingly important. For example, with the boom in search engines and bibliographic databases, the title and abstract are often the only two parts of a research paper that a potential reader can freely view, while access to the full paper may be subject to charges to the copyright owner~\cite{tullu2017writing}. In addition, when researchers conduct systematic investigations of related work, they have to spend time reading the full manuscripts if the corresponding abstracts are obscure, so they may abandon researching them~\cite{alspach2017writing,tullu2019writing}. In addition, during the blind review process, editors use abstracts to invite appropriate reviewers with expertise in the relevant field to evaluate papers~\cite{alspach2017writing,tullu2017writing}.

\par Concerns about the academic abstract writing skills of undergraduate and graduate students in higher education are well documented~\cite{chin2016investigating,article,tangpermpoon2008integrated}. From a faculty member's perspective, writing well is more than just following writing conventions. It also involves creative inspiration, problem-solving, reflection, and editing, culminating in a complete manuscript~\cite{kirkland1991maximizing,chin2016investigating}. From a student's perspective, writing an abstract can be a daunting task, both in terms of getting ideas on paper and mastering writing rules such as \textit{logic}, \textit{summary}, \textit{argument}, and \textit{grammar}~\cite{frey2003s,chin2016investigating}. To help students develop the abstract writing skills typically included in paper writing skills, institutions, such as universities, have conventionally recommended that students attend thematic writing training courses, such as scientific paper writing and biology essay writing, during which it is important for individual students to receive ongoing formative feedback~\cite{black2009developing}. However, the need to provide optimal formative feedback on individual abstract writing training in traditional large-scale lectures is often hampered by limited financial and pedagogical resources. One possible solution for providing individual feedback is to take advantage of recent advances in Natural Language Processing (NLP) and Machine learning (ML).

\par We systematically reviewed the literature on abstract writing in the field of educational technology following the rigorous approach suggested by Brocke et al. ~\cite{vom2015standing}. However, we found that the existing literature is under-researched in terms of \textbf{academic abstract writing} training. In contrast, a considerable number of tools have been developed to improve students' \textbf{summary writing} skills. It should be noted that abstracts and summaries are different\footnote{\url{https://smartleadershiphut.com/writing/abstract-vs-summary/}}\footnote{\url{https://www.scribbr.com/frequently-asked-questions/abstract-vs-summary}}\footnote{\url{https://www.mimjournal.com/post/main-differences-between-a-summary-and-an-abstract}}. While there are nuances to various accounts of the difference between an abstract and a summary, the general perception is that a summary of an entire article is a more detailed version of an abstract and that an abstract is usually written in the order of the content of a research paper, while a summary may focus on important aspects of the article\footnote{\url{https://smartleadershiphut.com/writing/abstract-vs-summary/}}\footnote{\url{https://www.scribbr.com/frequently-asked-questions/abstract-vs-summary/}}. Despite the differences, it is important to acknowledge that both contain important content and require students to have the ability to condense information.

\par We borrow the research ideas of summary writing training, from which we can conclude that the systems or methods proposed in computer-assisted summary writing training usually involve a three-stage cycle, namely: \textit{reading}, \textit{writing}, and \textit{feedback}. First, reading and understanding the main idea of the source text is critical. Previous work~\cite{sung2016effect} used concept maps to help students identify the main ideas and understand their hierarchy. While it may be suitable for general summary writing training, it is not appropriate in the scenario of academic abstract writing training because the concept maps in~\cite{sung2016effect} need to be generated by consultants and experts, which is too labor-intensive, especially when it comes to academic papers. In order to annotate concept maps for papers in different fields, a large number of experts are needed, as academic terminology and writing styles vary widely. Even in the same discipline, such as human-computer interaction (HCI), abstracts are written differently because of the different types of papers; for example, an application paper and a survey paper typically have different abstract writing styles. In terms of writing, existing summary training tools provide paradigms and summary strategies to instruct writing, which are good guides for abstract writing training. For the last stage, according to the literature review, there are four types of feedback, viz. \textit{providing scores}~\cite{crossley2019automated,chew2020enhancing,he2009automatic,wade2004summary}, \textit{peer review}~\cite{yang2016transforming}, \textit{section content coverage}~\cite{crossley2019automated,chew2020enhancing,wade2004summary} and \textit{summary writing strategy detection}~\cite{crossley2019automated,abdi2016automated,idris2011identifying}. However, for the same reasons as concept maps, the coverage of chapter content that requires instructor annotation does not apply to academic writing scenarios. To the best of our knowledge, there are no principles and proofs in the current literature on how to design automatically adaptive computer-assisted academic abstract writing tools to help a researcher learn abstract writing styles and patterns in his/her field.

\par To clarify the current status and main concerns of the abstract writing and training process for academic papers, we first systematically reviewed the literature in the field of pedagogy and educational technology~\cite{vom2015standing}. Then, we investigated the pain points of L2 (second language) junior researchers when writing abstracts through a formative study (a survey of $164$ students and semi-structured interviews with $11$ students). We aimed to address the three following research questions: 1) \textbf{RQ1: What are current student practices when writing abstracts?} 2) \textbf{RQ2: What are the specific challenges students have when writing abstracts?} and 3) \textbf{RQ3: What kind of support do students need when writing abstracts?} For \textbf{RQ1}, we learned that most L2 junior researchers write abstracts at least after completing the introduction part. For \textbf{RQ2}, we found that all the potential assistance we thought based on the literature review and summary writing tools were acknowledged by the participants. For \textbf{RQ3}, we extracted four main barriers that learners face when writing abstracts, namely: \textit{lack of skills in rephrasing content}, \textit{organization of ideas}, \textit{identification of main ideas}, and \textit{writing style recognition}. First, rephrasing is considered to be one of the core skills for paraphrasing key content, which is the essence of abstract writing~\cite{ashrafzadeh2015vocabulary}. However, students, especially L2 learners, resort to copying sentences from other parts of the paper rather than rewriting the main ideas in their own words~\cite{ashrafzadeh2015vocabulary,frey2003s}. Second, when it comes to organizing the ideas in each section of the paper, most junior students are not skilled at integrating them in a logical and cohesive manner while making the essay fluent and clear~\cite{ashrafzadeh2015vocabulary}. Third, despite the ability of novice researchers to identify the topic of the essay, secondary and irrelevant information remained easily incorporated, meaning they are deficient in grasping the complete hierarchy of ideas in the text~\cite{chin2016investigating}. Fourth, $73\%$ of students mentioned in their interviews that it would have been better to show the writing style or at least give them some hints. It can be quite time-consuming for them to align abstract ideas with lengthy original texts.

\par To address the above issues and fill the gap in abstract writing training, as well as to take advantage of recent advances in NLP technology, we propose a domain-oriented abstract writing training system \textit{ALens} (abbreviation for Abstract Lens), an adaptive learning tool that uses rhetorical structure parsing to identify main ideas, evaluates their abstracts from different linguistic features and uses visualization to analyze the writing patterns of reference abstracts (i.e., ground truth abstracts). Specifically, to address the first challenge and train users in their paraphrasing, we incorporate linguistic features, such as lexical and syntactic complexity, as assessment metrics. To address the idea organization problem, we run a re-trained sentence classification model that classifies abstract sentences into five genres (i.e. background, objective, method, result, and conclusion)~\cite{dos1996textual,cals2013effective,frey2003s,jalalian2012writing} and show the results in different colors. Considering the classification feedback, self-regulation~\cite{argote2000knowledge} regarding the organization of the abstract will be evoked, which will lead the user to discover which parts of the paper need to be included in specific areas and whether the ideas are expressed in a logical and cohesive order. To address the third challenge, we use discourse parsing with Rhetorical Structure Theory (RST)~\cite{heilman2015fast,mann1987rhetorical} to construct RST segments from the perspective of identifying logical relationships in the introduction. It separates sentence groups into RST trees with phrases on leaf nodes and logical relations on branches. RST uses rhetorical relations (e.g., elaboration, contrast, etc.) to depict the structure and logic of various parts of the text~\cite{mann1988rhetorical}. By parsing different paragraphs or the whole introduction, users can obtain the hierarchical structure of the text at different granularities and grasp the hierarchy of ideas in the text. Finally, to solve the last issue, following the approach utilized in the works about attention~\cite{vig2019multiscale,vig2019analyzing,bastings2020elephant,strobelt2018s}, \changed{we attempt to find relevant tokens in the generated abstract from the source text, apply semantic similarity to align the ideas between the reference abstract and the source text and reconstruct the style used by the authors in writing the reference abstract.} In addition, about 27\% told us in the interviews that when they have some ideas to write about but do not know where to start, they may get stuck. To facilitate the writing of the first draft, we embed a summary model as an option in the system ~\cite{guo2021longt5} to generate an initial draft as a prompt to start.

\par With the proposed research prototype, we further explored the
following two research questions: \textbf{RQ4: What is the technology acceptance level among junior researchers?} and \textbf{RQ5: How effective is \textit{ALens} in helping users write abstracts compared to the baseline system?} To answer these questions, we demonstrate the impact of \textit{ALens} on users' abstract writing skills by evaluating our system in two writing training scenarios. We quantitatively compare an abstract writing training method with our system. In a user study with $21$ students, the results show that with the help of \textit{ALens}, users could  organize their content in a more appropriate style when writing abstracts than the alternative tool. In addition, we measure the technology acceptance, user satisfaction, and engagement of both tools using the key constructs~\cite{venkatesh2003user,venkatesh2008technology}, and the results are encouraging, suggesting that \textit{ALens} can motivate students to learn abstract writing patterns in their own domain and to write abstracts in an appropriate style. Taken together, the main contributions of this work are:
\begin{compactitem}
\item We conduct a formative study to understand the problems encountered by \changed{L2 junior researchers} in the academic abstract writing process.
\item We build \textit{ALens}, an automatic feedback learning tool that first incorporates visualization and interactive features into academic abstract writing training.
\item We show the effectiveness of \textit{ALens} by comparing it with an alternative abstract writing training tool.
\end{compactitem}

\section{Related Work}
\par The literature that overlaps with this work can be grouped into four categories, namely, \textit{technology-mediated summary writing assistance}, \textit{summary evaluation metrics}, \textit{NLP models in the summary task}, and \textit{self-regulated learning}.
% which is the potential theory for our main assumptions.}

\subsection{Technology-Mediated Summary Writing Assistance}
\par We systematically reviewed the literature on abstract writing in the field of educational technology following the rigorous approach suggested by Brocke et al.~\cite{vom2015standing}. \changed{However, although several tools have been developed to improve students' summary writing skills over the past decade, very little literature has focused on the development of learning tools for abstract writing.} The main difference between an abstract and a summary of a whole article is the length and purpose\footnote{\url{https://smartleadershiphut.com/writing/abstract-vs-summary/}}\footnote{\url{https://www.scribbr.com/frequently-asked-questions/abstract-vs-summary}}\footnote{\url{https://www.mimjournal.com/post/main-differences-between-a-summary-and-an-abstract}}. Abstracts usually follow the empirical order of content as specified by the journal or association and cover the main aspects of the research paper. Summaries may not follow specific guidelines, emphasizing certain important aspects of the paper and providing more details than the abstract. Despite the differences, it has to be acknowledged that both are abbreviated versions of the paper that contain important content and require the ability to understand, express, synthesize and paraphrase~\cite{spirgel2016does,brown1981learning,brown1983macrorules}. For example, in a study of computer-assisted summary writing training, the \textbf{concept map}~\cite{chew2020enhancing} arranges concepts in the text in layers, with general concepts at a shallower level and specific concepts at a deeper level. It attempts to facilitate students' identification of the main ideas and understanding of the corresponding supporting ideas. Several studies have proposed methods that identify the \textbf{summarization strategies}, including deletion, sentence combination, and paraphrasing used by students to help assess teachers' summarization processes and to target them during training. \textbf{Worked examples}~\cite{chew2020enhancing,wade2004summary,he2009automatic} are exemplars with worked-out steps and predetermined questions and are often used as guides to help students learn to read the original text and summarization strategies. In addition, by comparing multiple worked examples, students gain the ability to identify patterns of relevant and irrelevant information~\cite{chew2020enhancing}. The \textbf{Computer-Supported Collaborative Learning (CSCL) approach}~\cite{yang2016transforming} is embedded in the summary writing training system, and students receive peer feedback through online conversations and interactions. As they digest peer feedback, students reflect on their summarization process and make further revisions~\cite{hovardas2014peer}.

\par However, the above approach cannot be directly applied to academic abstract writing training. Specifically, concept maps and worked examples are carefully prepared by instructors and need to be annotated article by article due to the different topics and progression of the articles. In other words, when it comes to academic papers, the workload of instructors in generating concept maps and worked examples can be very high. To fill the gap in abstract writing training and to take advantage of recent advances in NLP technology, we use rhetorical structure parsing to identify main ideas, evaluate abstracts in terms of different linguistic features, and use visualization to analyze the writing patterns of reference abstracts.

\subsection{NLP Models in Summary Tasks}
\par The text processing models behind text summarization tools can be broadly classified into two categories, namely \textit{extraction} and \textit{abstraction}~\cite{el2021automatic}. Extractive approaches~\cite{jia2020neural,liu2019text,zhong2020extractive,mihalcea2004textrank,nallapati2017summarunner,see2017get} \textbf{copy salient phrases and sentences from the text and merge them to create summaries}~\cite{nallapati2017summarunner,zhong2020extractive}, thus ensuring that the summaries are factually consistent with the source text~\cite{cao2018faithful}. However, the extraction paradigm is often criticized for being logically inconsistent with the input text~\cite{see2017get,see2021neural}. Abstraction methods~\cite{wu2021r,zhang2020pegasus,guo2021longt5,lewis2019bart} \textbf{rearrange the language in the text and add new words or phrases to the abstract as needed}~\cite{gupta2019abstractive}. Since state-of-the-art abstraction methods perform well in generating fluent human-like summaries~\cite{zhang2020pegasus}, in our work we embed the abstraction summarization models into our system~\cite{guo2021longt5} as an option for generating the initial manuscript to prompt the user to start.

\subsection{Summary Evaluation Metrics}
\par In the learning process, it is important to provide individual and adaptive feedback~\cite{black2009developing}, and the same is true for abstract writing training. We consider assessment methods widely used in summary writing training as a potential approach to abstract writing training. For example, assessment scores are a typical method of providing formative feedback in computer-assisted abstract writing training~\cite{sung2016effect,wade2004summary,chew2020enhancing}, and there are three types of scores, i.e., content coverage scores~\cite{wade2004summary,chew2020enhancing}, scores given by mathematical methods~\cite{crossley2019automated,lin2004rouge,zhang2019bertscore,kondrak2005n} and scores predicted by pre-trained language models~\cite{xenouleas2019sumqe,bohm2019better,louis2013automatically}. Specifically, content coverage scores are calculated automatically to measure the degree of coverage of each content in the summary. Although calculated automatically, the exact content to be measured is specified by the instructor on an article-by-article basis. However, this is clearly not appropriate for academic abstract writing, as the differences in disciplines, fields, and paper types result in a significant amount of work for instructors to develop content criteria for each type of article. Instead, mathematical methods and deep learning approaches are the most suitable candidates. Although pre-trained deep learning language models can achieve a high degree of agreement with human estimates, their high performance is highly dependent on the availability of relevant datasets. Due to the unavailability of high-quality datasets, we turn to mathematical methods. Specifically, three methods are commonly used: metrics in ML~\cite{lin2004rouge,zhang2019bertscore,papineni2002bleu}, latent semantic analysis (LSA)~\cite{landauer2013handbook} and linguistic features~\cite{kyle2018tool,kyle2016measuring,crossley2016tool}. Metrics in ML, e.g., \textit{ROUGE}~\cite{lin2004rouge}, \textit{BERTScore}~\cite{zhang2019bertscore} and \textit{Bleu}~\cite{papineni2002bleu} and LSA all assess the quality of a summary based on semantic overlap with the reference or source text, thus giving an overall score for the summary. However, this single score is not an appropriate feedback~\cite{weigle2002assessing}, and it does not reveal the gap between what people understand and what they should understand~\cite{sadler1989formative}. Therefore, we rate the summaries using different scoring criteria (e.g. lexical complexity and cohesion) based on linguistic features, which is considered more appropriate because it captures different aspects of the summaries and thus provides more informative and instructive feedback~\cite{bernhardt2010understanding}.

\subsection{Self-regulated Learning}
\par It has been hypothesized that providing students with feedback about their writing abilities will enhance their learning experience and facilitate the writing of high-quality summaries~\cite{zimmerman2001self}. In order to achieve self-regulated learning, providing students with formative feedback as well as setting goals is essential~\cite{bandura1997social}. It has been argued that in order for feedback systems to be effective, learners must be provided with goals, their progress tracked, and actions identified to help them achieve those goals~\cite{hattie2007power}. However, individuals are unable to track their own progress in the work~\cite{bjork2013self}. Using targeted assessment and feedback is a good way to enhance the learning process~\cite{roediger2006test}. When students are given feedback on their abilities throughout the intervention, it can increase their chances of achieving better short-term outcomes on specific learning tasks~\cite{hattie2007power,roediger2006test,artemeva2008toward}. In this work, we provide students with user-centered adaptive feedback about their abstracts to determine if they can write and improve organized abstracts.

\section{Formative Study}
\par Based on the similarities between summary writing and abstract writing in terms of the writing process and required competencies, and in order to fill the gap in abstract writing training tools, we adopt a top-down approach, first distilling possible meta-requirements from the existing literature on summary writing training tools. With this goal in mind, we first selected $27$ papers discussing summary writing training for meticulous analysis, from which we distilled the closed loop of summary writing learning. In addition, because abstract writing aids span the fields of education, psychology, and computer science, we focused on literature in these categories. On this basis, additional $32$ related papers were selected to further analyze and understand established pedagogical theories in writing~\cite{belcher1995academic} and metacognition~\cite{livingston2003metacognition} in the learning process, which is considered a meta-need for adaptive learning tools.

\par Next, in order to derive the user requirements for the academic abstract writing training system, we first need to understand the problems that students encounter in the academic writing process. Therefore, we design and examine the following three research questions: 1) \textbf{RQ1: What are current student practices when writing abstracts?} 2) \textbf{RQ2: What are the specific challenges students have when writing abstracts?} and 3) \textbf{RQ3: What kind of support do students need when writing abstracts?}

\subsection{Survey Study}
\subsubsection{Survey Protocol}
\par The survey was administered on the Microsoft Forms online platform. The survey questions included abstract writing practices, difficulties in abstract writing, assistance needed when writing abstracts, and demographic questions. The survey contained textual questions and ranking questions about the academic paper writing process. It also contained questions about students' challenges. 5-point Likert scale questions are used to measure students' attitudes toward several potential types of assistance. The survey also contained open-ended questions about student requests. At the end of the survey, respondents were allowed to leave their contact information if they wished to be interviewed for follow-up.

\subsubsection{Respondents and Recruitment}
\par We recruited $164$ respondents ($54$ female, $106$ male, and $4$ prefer not to specify) between the ages of $19$ and $36$ (B.S.: $105$, M.S.: $42$, Ph.D.: $13$, Others: $4$) via advertised posts in online university communities. Of all respondents, $125$ with academic writing experience answered Q$1$ -- Q$3$ and the other $39$ answered Q$2$ -- Q$3$.

\begin{table*}[h]
\centering
\begin{tabular}{ccccc|ccc}
\bottomrule
\multicolumn{5}{c|}{\textbf{Had academic writing experience}} & \textbf{Potential assistance} & \textbf{M} & \textbf{SD} \\ \cline{6-8} 
\multicolumn{4}{c}{\textbf{Yes}} & \textbf{No} & Online revision & 4.24 & 0.88  \\
\multicolumn{4}{c}{125/164} & 39/164 & Present key information of intro & 4.22 & 0.69  \\ \cline{1-5}
\multicolumn{4}{c|}{\textbf{Used writing aids}} & \multirow{6}{*}{/} & Word count & 4.21 & 0.35 \\
\multicolumn{2}{c}{\textbf{Yes}} & \multicolumn{2}{c|}{\textbf{No}} &  & Present abstract structure & 4.17 & 0.84 \\
\multicolumn{2}{c}{67/125} & \multicolumn{2}{c|}{58/125} &  & Abstract writing style recognition & 4.10 & 0.88 \\ \cline{1-4}
\multicolumn{4}{c|}{\textbf{Wrote abstract referring to}} &  & Abstract evaluation & 4.09 & 0.86 \\
\textbf{Intro} & \textbf{Body} & \textbf{Full text} & \multicolumn{1}{c|}{\textbf{Others}} &  & Support intro annotation & 4.08 & 0.83 \\
29/125 & 45/125 & 47/125 & \multicolumn{1}{c|}{4/125} &  & Present logic relation of sentences & 4.06 & 0.88 \\ \cline{1-5}
\multicolumn{3}{c}{\multirow{2}{*}{\textbf{Total respondents number}}} & \multicolumn{2}{c|}{\multirow{2}{*}{164}} & Present intro structure & 4.06 & 0.88 \\
\multicolumn{3}{c}{} & \multicolumn{2}{c|}{} & Instructional feedback & 4.03 & 0.91  \\ \bottomrule
\end{tabular}
\vspace{5pt}
\caption{\changed{Results of the survey. On the left is the distribution of the number of distinct respondents; on the right is a tally of 5-point Likert scale questions about potential help ($1$ -- $5$: very unhelpful -- very helpful).}}
\label{tab:survey_result}
\vspace{-6mm}
\end{table*}

\subsection{Interview Study}
\subsubsection{Interviewees}
\par To gain more insight into the challenges and requirements of students when writing their abstracts, we further contacted $11$ students ($3$ female, $8$ male; $10$ graduate students, $1$ PhD. student) who left their email addresses in their questionnaire. Their ages ranged from $20$ to $26$ (mean age = $23.09$, SD = $1.81$).

\subsubsection{Interview Protocol and Analysis Method}
\par We conducted remote semi-structured interviews using an online communication tool and audio-recorded the interviews with consent. The interviews consisted of three main sections: (1) how students typically write abstracts; (2) the challenges in writing abstracts; and (3) what features students need. \changed{To analyze the challenges students encounter when writing abstracts for academic papers, we followed an iterative coding process~\cite{hruschka2004reliability} for thematic analysis. For each question, one author open-coded the responses to identify the categories that appeared and developed a codebook. We noted that a single response may include multiple categories. Therefore, we treated each category as binary. For each response, we labeled whether each category was present or absent. Two coders coded all responses independently. They then discussed inconsistencies, refined
the code definitions, and independently re-coded the responses based on the new definitions. They iteratively coded the responses until they reached a Cohen's kappa above $0.7$ for all categories. Finally, we came up with five subcodes for the students' challenges.}

\subsection{Findings and Design Requirements} 
\par For RQ$1$, We recapitulated the following findings from the survey and interview results. Literature~\cite{wallwork2016english,khadilkar2018art} and reports\footnote{\url{https://writingcenter.gmu.edu/writing-resources/different-genres/writing-an-abstract}} indicate that ``\textit{...the Abstract must be written after completing the entire manuscript. Ensure that important points made in the main manuscript are included in the abstract...}''. We also found that most learners (survey: $139/164$, interview: $10/11$) wrote the abstract after writing the introduction or body of the paper. They usually first determined the structure of the abstract, i.e., what sections need to be included and which are more important. Then they wrote down each part purposefully. And most learners (survey: $119/164$, interview: $8/11$) would refer to the introduction or body of the paper to ensure consistency. For example, after reading the introduction, some of them would extract important sentences or paragraphs by highlighting these texts and reorganizing these texts into an abstract. More than half of them would use writing aids such as Grammarly\footnote{\url{https://app.grammarly.com}}. For RQ2, The detailed results of the survey are shown in \autoref{tab:survey_result}. The result shows that all the potential assistance we refined from the summary writing training tool and  pedagogical theories are verified by our respondents. The main findings for RQ3 are summarized in \autoref{challenges_of_academic_abstract_writing}.

% \usepackage{multirow}

\subsubsection{Challenges of Academic Abstract Writing}
\label{challenges_of_academic_abstract_writing}
\par We combined our survey research and interview study to present the following five challenges.
\par \textbf{C1: Lack of skills in rephrasing content \changed{(N=$7/11$)}.} Sometimes students tend to rewrite key sentences in the introduction, especially when writing background and conclusion sentences (P1, P3 -- P7). However, rephrasing content is sometimes tricky because students cannot directly use sentences from the introduction. Current summarization techniques produce wording that is still too close to the original text, which does not help solve this problem. ``\textit{How to express the same meaning precisely in a new way is sometimes an annoying problem (P2, male, age=24).''}\

\par \textbf{C2: Identification and organization of ideas \changed{(N=$9/11$)}.} \changed{On the one hand, according to the survey results, almost half of the respondents \changed{($47.6$\% of the 164 participants)} answered that it was quite difficult to \textit{summarize all the key points in a limited space}, or to \textit{write them concisely enough.} }On the other hand, having a good insight into the logic of the introduction is crucial for students to write excellent abstracts. According to the survey results, most students \changed{($82.96$\% of 164 participants)} wrote their abstracts after writing the main body of the paper. However, students may forget the logical flow of the introduction after writing the main body of the paper, especially sections with complex logical relationships. For example, ``\textit{in my field, the prior experiments section in the introduction includes too many experimental methods. When writing the abstract, I always need to figure out again how they relate to each other (P9, male, age=$21$).}''

\par \textbf{C3: Recognition of writing pattern and style \changed{(N=$8/11$)}.} Abstract writing is often field-oriented, as different disciplines and different types of papers, and different journals and conferences generally differ in style and writing patterns. The survey found that \changed{$53.0$\% of 164 respondents} found it time-consuming to master the style abstracts and find their regularity by perusing articles in the field, and 65.9\% thought it would be better if they were shown the regularity.

\par \textbf{C4: Requirement for the first draft.} \changed{$3$ out of $11$ students} responded in the formative interview that they did not know where to start writing. ``\textit{I'm used to revising from other people's drafts and I can't start from a completely blank space (P7, female, age=22).''}

\subsubsection{Design Requirements}
\par Based on the identified challenges in academic paper abstract writing and users' expectations for satisfactory assistance results and comprehensive functionality, we derive the following design requirements of an adaptive abstract writing training tool.

\par \textbf{R1: Provide assistance on rephrasing.} Through the formative study, we found that how to rewrite key sentences and other information extracted from the introduction is a great challenge for learners \textbf{(C$1$)}. To address this issue, we should provide guidance on representing information in another way, such as sentence transformation and phrase substitution based on self-regulated learning theory~\cite{bandura1997social}.

% test
% In \nameref{label:text} we have
% \labelText{This \textsc{is} a text that is also a tag}{label:text}

\par \textbf{R2: Help learners better understand the introduction \changed{and organize the main ideas}.} From the previous formative study, we found that many learners encountered difficulties in selecting core information in a limited space, i.e., the problem of main idea identification \textbf{(C$2$)}. To effectively address this problem, learners' mastery of the structure and content of the introduction is exceptionally demanding. On the one hand, the introduction is a distillation of the main text, and the relationship between some sentences is difficult to grasp. Therefore, we should provide guidance on identifying the logical relationships of sentences in the introduction. On the other hand, the introduction is long and requires extra time to reread because the content is forgotten. We should also help learners quickly review the structure and content of the introduction. After understanding the introduction, some learners still have difficulty organizing these key elements fluently \textbf{(C$2$)}. We should help them to have a better understanding of the information from a new perspective and help them to organize the main ideas in a rational way.

\par \textbf{R3: Assist learners in understanding domain-specific abstract styles.} As shown in \autoref{tab:survey_result}, most participants indicated that knowledge about how abstracts are written was very useful to them (Mean = $4.10$, SD = $0.88$) \textbf{(C$4$)}. The underlying style can guide learners to write abstracts that are more accurate in content and organization. Therefore, in addition to directly presenting the reference abstract, we should also demonstrate its style in a clear and intuitive manner \textbf{(C$3$)}. \looseness=-1

\par \textbf{R4: Prepare for the first draft.} In the semi-structured interviews, $3$ students mentioned the difficulty of writing abstracts from scratch. Despite the relatively low rate, we observed the rise of text summarization platforms such as \textit{TLDR this}\footnote{\url{https://tldrthis.com/}}, \textit{Resoomer}\footnote{\url{https://resoomer.com/en/}}, and \textit{Wordtune Read}\footnote{\url{https://app.wordtune.com/read}}. Therefore, we believe that there is a trend to harness the power of NLP techniques to facilitate abstract writing, for example, to generate the first draft version \textbf{(C$4$)}.

\par \textbf{R5: Easy to access and use.} From the survey, some potential users expressed concerns about the complexity and difficulty of using the features of the general academic abstract writing assistance system. Therefore, we had to ensure that the tool would not become burdensome and responsive to users while providing practical features to address the above challenges. For example, users did not need to install additional software or hardware, using the typical writing assistance platform interface design familiar to learners. \changed{Despite the above requirements, basic functionality is also highly valued, as shown in \autoref{tab:survey_result}}. Therefore, the tool also needs to have the basic features of a writing aid to ensure effectiveness, such as online revision and word count functions.

\section{Design of the Abstract Writing Training System}
\subsection{Approach Overview}
\begin{figure*}[h]
 \centering % avoid the use of \begin{center}...\end{center} and use \centering instead (more compact)
 \includegraphics[width=\textwidth]{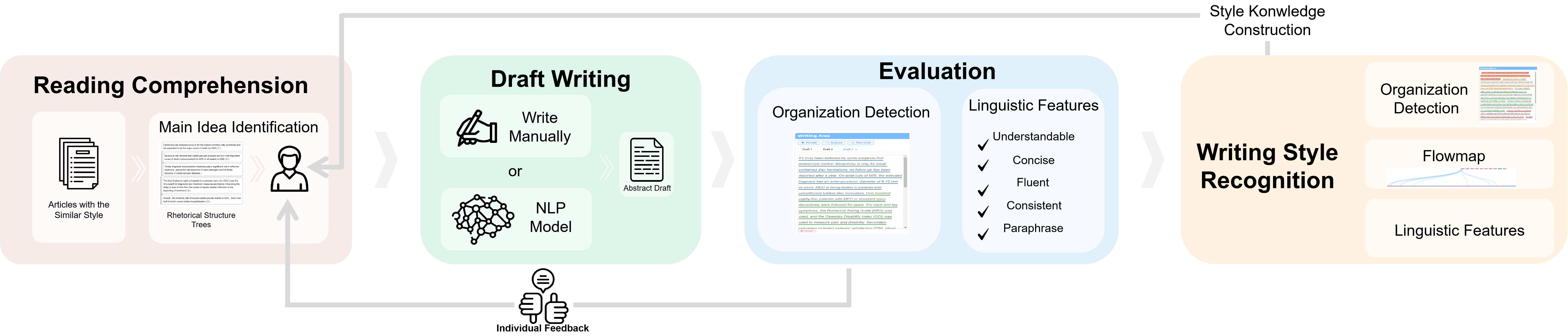}
  \vspace{-6mm}
 \caption{Pipeline of \textit{ALens}: (1) read and comprehend the source text (2) generate the first abstract draft for later revision; (3) refine the abstract draft considering organization, main idea, and lexicon; evaluate the results based on semantic analysis; (4) recognize and learn the style of the reference abstract.}
 \label{fig:pipeline}
   \vspace{-3mm}
\end{figure*}

\par Based on the requirements derived from the formative study, we design an abstract writing training process and incorporate it into a web-based writing assistance platform named \textit{ALens}. It facilitates users to quickly grasp the main ideas of an essay, optionally write using a summarization model from NLP, recognize their deficiencies in abstract writing, and gain knowledge about style in specific scenarios. To support the writing process in a convenient and user-friendly manner \textbf{(R5)}, visualization and interaction are integrated into \textit{ALens} to cater to the mental habits of users with different granularity requirements. \textit{ALens} consists of a \textit{Rhetorical Structure View}, a \textit{Writing Area}, an \textit{Evaluation Dashboard}, and a \textit{Reference Abstract with a Flow Map}. \autoref{fig:pipeline} describes the general stages of the designed abstract training pipeline. First, the user can select an article to be learned for abstract writing and upload it. Subsequently, the rhetorical structure of the original article is analyzed to help the user quickly identify the main ideas in terms of logical structure \textbf{(R2)}. Then, the user can choose to write the first draft from scratch or with the help of a summarization model \textbf{(R4)}. Given the lack of content organization, the sentences in the abstract are divided into several types (e.g. background and conclusion)~\cite{andrade2011write,cohan2018discourse, meng2021bringing}, and the completeness of the abstract is checked against the domain of the paper, i.e., whether the first draft properly covers and arranges the domain typical of the abstract's required information and guide users to reflect on them \textbf{(R2)}. Meanwhile, \textit{ALens} can automatically analyze the linguistic features of the abstract to check whether it is comprehensible, concise, fluent, and consistent with the source text \textbf{(R1)}. In addition, paraphrase detection is applied to the feedback to guide the user to rephrase sentences instead of copying them. Finally, users can check the writing style of the reference abstract and analyze its linguistic features and organization. Specifically, a flow map is used to align ideas in the reference abstract with the source text, that is, to find the most relevant content from the source text. By comparing these features of different articles with similar academic domain ``styles'', users are expected to discover writing patterns and learn writing styles \textbf{(R3)}.

\subsection{NLP Pipeline}
\par The back-end engine of \textit{ALens} first help users to identify ideas by parsing an article into a rhetorical tree, and then supports the production of the initial draft for later revision. The sentences in the abstract are then divided into several genres to detect the organization of the abstract. At the same time, the abstract is evaluated by different linguistic features, providing personal feedback to stimulate revision according to a self-regulated learning theory~\cite{bandura1997social}.

\subsubsection{Main Idea Identification}
\par
\begin{figure}[h]
 \centering
 \includegraphics[width=0.3\textwidth]{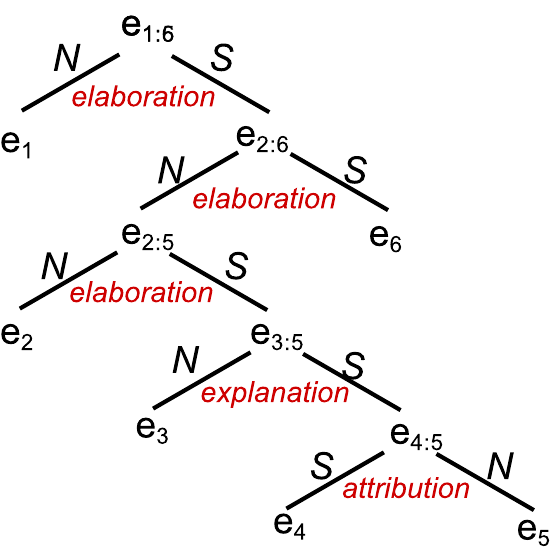}
 \vspace{-2mm}
 \caption{\changed{$e_1$[ Compare the past eight five-year plans with actual appropriations. ] $e_2$[ The Pentagon's strategists produce budgets ] $e_3$[ that simply cannot be executed ] $e_4$[because they assume] $e_5$[ a defense strategy depends only on goals and threats. ] $e_6$[ Strategy, however, is about possibilities, not hopes and dreams. ] 
 \newline
 An example of an RST structure tree from the RST discourse tree bank\cite{lynnmarcu2002}. $e_i$, $e_{j:k}$, $N$, and $S$ denote basic discourse units, spans, nucleus, and satellite, respectively.}}
 \label{fig:rstTree}
  \vspace{-3mm}
\end{figure}

 To enable learners to quickly grasp the hierarchical structure of a text and avoid wasting time by repeatedly reviewing the introduction when writing an abstract, we provide a rhetorical structure parsing for each paragraph. In particular, we provide rhetorical relationship recognition for any text with a continuous span. For example, after inputting a text containing three sentences with six elementary discourse units (EDUs: tokens of adjacent text spans, roughly analogous to independent phrases) (e$_1$ -- e$_6$), the model can output relations for any continuous EDUs. As shown in \autoref{fig:rstTree}, \changed{e$_2$~ -- ~e$_6$ is an elaboration of e$_1$, e$_4$ -- e$_5$ is an explanation of e$_3$, e$_5$ is attributed to e$_4$, and so on.} All relations and their hierarchy build the structure of the whole text and give a deeper understanding of the text structure.

\par The identification of rhetorical relations consists of two parts: 1) text segmentation \changed{(TS)} and 2) relation identification ~\changed{(RI)}. For text segmentation, we need to segment the input text into EDUs. Each sentence consists of several EDUs. We retrained the model proposed by Heilman et al.~\cite{heilman2015fast} to operate the segmentation task with high accuracy. It uses a conditional random field \changed{(CRF)} model with $l_2$ regularization\cite{fields2001probabilistic}. For each token in a sentence, it predicts the whether it is the beginning of a new EDU or not. \changed{The CRF regularization parameter was $64.0$, adjusted by grid search using a grid of powers of $2$ between $1/64$ to $64$. Compared to the human agreement (HA)\cite{bach2012reranking}, the percentages of precision, recall and F1 score were $90.2$(TS)/$98.5$(HA), $83.5$(TS)/$98.2$(HA), and $86.7$(TS)/$98.3$(HA), respectively.} The second relation recognition component was modeled as a classification problem, i.e., classifying the relation of two consecutive EDUs into 16-tuple types such as elaboration, contrast, and joint. For this task, we utilized the ZPar model~\cite{zhang2011syntactic} as the relationship parser, which predicts the rhetorical relationships across the text at different levels of granularity. The output of the model is the relationship of EDUs, i.e., the phraseological relationships in a sentence, which is not valuable for abstract writing. Therefore, we modified and retrained the ZPar model to predict rhetorical relations between adjacent sentences at a more appropriate level of granularity. \changed{The parser was estimated using multiclass logistic regression with an $l_1$ penalty. $l_1$ was adjusted after the grid search and finally set to $0.25$. Compared to the human agreement (HA)\cite{ji2014representation}, the parser performs $83.5$(RI)/$88.7$(HA), $68.1$(RI)/$77.7$(HA), and $55.1$(RI)/$65.8$(HA) in terms of span, kernel, and relationship, respectively. The parser model uses a shift-reduced algorithm with a time complexity of $O(n)$, where $n$ is the number of EDUs.}

\subsubsection{Summarization Assistance}
\par It is important to note that providing a relatively reliable first draft of the abstract is an important foundation for the later revision process. However, a requirement for developing NLP models with the ability to summarize research papers is the availability of relevant datasets. We reviewed the literature on corpora and found that the \textit{arXiv} and \textit{PubMed} datasets released in~\cite{cohan2018discourse} met our requirements. \textit{PubMed} contains $119k$ article body and abstract pairs of the biomedical literature, while \textit{arXiv} contains $203k$ pairs of articles on different topics. Since different subjects have different terminologies and writing styles, we decided to construct a subset called \textit{arXiv-cs} that contains only computer science (cs) articles by iterating through the abstracts in \textit{arXiv} and match them in \textit{arXiv-dataset} released in~\cite{clement2019arxiv}, which hosts $1.5M$ metadata of preprinted articles in physics, mathematics and computer science from $1991$ to $2019$ to determine whether they belong to the computer science domain.

\begin{table*}[h]
\begin{tabular}{@{}ccccccccc@{}}
\toprule
     & \multicolumn{4}{c}{PubMed}               & \multicolumn{4}{c}{arXiv-cs}                \\ \midrule
     & Rouge-1 & Rouge-2 & Rouge-L & Rouge-Lsum & Rouge-1 & Rouge-2 & Rouge-L & Rouge-Lsum \\

   & 47.34   & 22.53   & 28.74   & 42.79      & 46.91   & 19.67   & 27.41   & 41.87      \\ \bottomrule
\end{tabular}
\vspace{5pt}
\caption{Results of fine-tuning \textit{LongT5} on two datasets in biomedicine and computer science.}
\vspace{-6mm}
\label{tab:finetune_result}
\end{table*}

\begin{table*}
	\centering
	\resizebox{\textwidth}{45mm}{
	\begin{tabular}{lrrr}
\toprule
Facets of Quality                                       &Linguistic Features                           &Definition \\ \midrule
\multirow{4}{*}{Understandability} & Frequency (COCA spoken, All Words)            &\makecell[c]{The sum of frequency scores of all words occurs in COCA spoken corpus \\divided by number of words in text with frequency score}     \\
                                  & Frequency (COCA spoken, Function Words)        &\makecell[c]{The sum of frequency scores of function words occurs in COCA spoken corpus\\ divided by number of words in text with frequency score}            \\
                                & Frequency (SUBTLEXus, Content Words)           &\makecell[c]{The sum of frequency scores of content words occurs in SUBTLEXus corpus\\ divided by number of words in text with frequency score}       \\
                               & Frequency (SUBTLEXus, All Words)                &\makecell[c]{The sum of frequency scores of all words occurs in SUBTLEXus corpus\\ divided by number of words in text with frequency score} \\
\midrule Consistency                                            & Source similarity (ROUGE-3)           &\makecell[c]{Similarity to the source, calculated by ROUGE-3}        \\
\midrule \multirow{3}{*}{Fluency}                               & Adjacent sentence similarity (word2vec)  &\makecell[c]{Similarity between two adjacent sentences, calculated by word2vec}       \\
                                                       & Repeated content lemmas and pronouns     &\makecell[c]{Number of repeated content and third person pronouns divided by number of words}      \\
                                                       & Binary adjacent sentence overlap (Function Words)          &\makecell[c]{Number of overlapping function words in two adjacent sentences}  \\
\midrule \multirow{11}{*}{Diversity}                            & Type-token ratio (All Words)                      &\makecell[c]{The number of unique words (types)\\ divided by the total number of words (tokens) in a given segment of language }   \\
                                                       & MATTR (Function Words)                                     & \makecell[c]{Moving average type token ratio for function words (50-word window)}\\
                                                       & Number of Content Words tokens                             &\makecell[c]{Number of Content Words tokens}\\
                                                       & MTLD (Function Words)                                    &\makecell[c]{Average number of function word tokens it takes\\ to reach a given TTR value (.720)}  \\
                                                       & MTLD (All Words)                                     &\makecell[c]{Average number of tokens it takes\\ to reach a given TTR value (.720)} \\
                                                       & MTLD (Content Words)                                      &\makecell[c]{Average number of content word tokens it takes\\ to reach a given TTR value (.720)}\\
                                                       & Lexical density (Percentage of Content Words)            & \makecell[c]{Percentage of Content Words in the text}\\
                                                       & Type-token ratio (Content Words)                          &\makecell[c]{The number of unique content words (types)\\ divided by the total number of words (tokens) in a given segment of language}\\
                                                       & SD of dependents per nominal subject           & \makecell[c]{Standard deviation  of dependents per nominal subject}\\
                                                       & SD of dependents per clause                   &\makecell[c]{Standard deviation  of dependents per clause} \\
                                                       & SD of dependents per object of the preposition &\makecell[c]{Standard deviation  of dependents per object of the preposition}\\
\midrule \multirow{2}{*}{Conciseness}                           & Mean length of sentence                       & \makecell[c]{Mean length of sentence}  \\
                                                       & Mean length of clause                       & \makecell[c]{Mean length of clause}     \\
                                                       & Word counts                       & \makecell[c]{Word counts} \\\bottomrule
	\end{tabular}}
  \vspace{2mm}
\caption{The table shows the five facets of abstract quality and the corresponding linguistic features used to calculate each facet. For specific definitions of linguistic features, please refer to~\cite{kyle2018tool,kyle2016measuring,crossley2016tool,kyle2021assessing}.}
\label{tab:facets}
\vspace{-3mm}
\end{table*}

\par Based on these two datasets (i.e.,\textit{arXiv-cs} and \textit{PubMed} datasets consisting of article body and abstract pairs), we developed the following fine-tuning scheme. Since the two datasets belong to the biomedicine and computer science domains, respectively, we developed an abstraction summarization model based on \textit{LongT5-large}\footnote{\url{https://github.com/google-research/longt5}}~\cite{guo2021longt5} for each domain. The training parameters follow those described in the original publication and GitHub; in addition, the maximum input token and maximum output token are set to $(4096, 512)$ on \textit{PubMed} and $(16384, 512)$ on \textit{arXiv-cs}, and the model performance for both fine-tunings is shown in \autoref{tab:finetune_result}. \changed{Note that Rouge (Recall-Oriented Understudy for Gisting Evaluation)~\cite{lin2004rouge} is a set of metrics for evaluating automated abstracts and machine translations. It measures the ``similarity'' between an automatically generated abstract or translation and a reference abstract by comparing it with a set of reference abstracts (usually manually generated) and calculating the corresponding score.} The results show that these models can be used to generate a relatively reliable abstract for later revisions. \changed{Please refer to \autoref{tab:hyperparams_summarization} in \autoref{Hyperparameters  of NLP models} for the hyperparameters we set in this work for details.}

\subsubsection{Organization Detection}

\par To facilitate the organization of the ideas distilled from the source text, we detected the organization in the abstract, thus inducing the user to think about the coverage and arrangement of the content. The organization detection task is considered as a multi-class sentence classification task, where each sentence in the abstract is classified into five types, i.e., \textit{background}, \textit{objectives}, \textit{methods}, \textit{results} and \textit{conclusions}~\cite{dos1996textual,cals2013effective,frey2003s,jalalian2012writing}. As for our choice of this classification scheme, the reason is the lack of domain-specific and annotated sentence classification datasets. For the sentence classification task, we found two datasets, \textit{PubMed 200k RCT}~\cite{dernoncourt2017pubmed}, containing $200k$ of type and abstract sentence pairs and \textit{CSAbatract}~\cite{Cohan2019EMNLP}, containing $2k$ of pairs corresponding to the biomedicine and computer science domains, respectively. The goal of the classification model is to provide accurate classification to identify sentence intent in the abstract, which can be used to assess the organization of the draft and thus induce self-reflection on how to improve the coverage and arrangement of the content. Following the \textit{BERT-base-uncased}\footnote{\url{https://github.com/google-research/bert}}, a model pretrained on \textit{BookCorpus} and \textit{English Wikipedia}\footnote{\url{https://en.wikipedia.org/wiki/English\_Wikipedia}} proposed in~\cite{devlin2018bert}, \changed{we fine-tuned on these two datasets separately and trained the model with different hyperparameters.} For the sentence classification task on \textit{PubMed 200k RCT}, the F1-score of the model is $83.59\%$, while for the task on \textit{CSAbstract}, the F1-score of the model is $86.37\%$. These results show that we can embed the BERT model into our system to provide users with organizational analysis. \changed{Please refer to \autoref{tab:hyperparams_classification} in \autoref{Hyperparameters of NLP models} for the hyperparameters we set in this work for details.}

\subsubsection{Evaluation Metrics}
\par As with abstract writing training, it is critical to provide individual and adaptive feedback during the learning process~\cite{black2009developing}. The summaries and abstracts synthesize the main ideas of the text and require the ability to understand, express, synthesize, and paraphrase~\cite{spirgel2016does,brown1981learning,brown1983macrorules}. Abstract writing training may benefit from the evaluation methods used in summary writing training. Computer-assisted summary writing training typically provides formative feedback in the form of assessment scores. However, this single score is not an appropriate feedback~\cite{weigle2002assessing}, so we tend to score abstracts using different scoring criteria. \changed{As a writing task, the scoring rubric for abstract writing should first focus on the scoring dimensions of general writing, i.e., content, content organization, and expression~\cite{guo2013predicting}.} In addition, from the literature~\cite{andrade2011write,vasilyev2020fill,vasilyev2020human} and the writing instructions on the website\footnote{\url{https://classroom.synonym.com/list-abstract-qualities-8671549.html}}\footnote{\url{https://www.brandeis.edu/writing-program/resources/students/handouts/features-of-a-good-abstract-handout.pdf}}\footnote{\url{https://www.abstractscorecard.com/uploads/cfp2/images/Abstract\_Quality\_Standards\_Guidelines\_13.pdf}}, we concluded that a good abstract should be comprehensible, concise, and fluent. In addition, it should be consistent with the source text and use words and phrases that are different from the source text. Linguistic features are considered more appropriate because they capture different aspects of the abstract and thus provide more informative and instructive feedback~\cite{bernhardt2010understanding}. Therefore, we use different linguistic features to compute the five aspects. First, we selected $21$ linguistic features that were shown to be related to the quality of abstracts to calculate their quality based on~\cite{crossley2019automated}. Then, we clustered the linguistic features to measure the five aspects of abstracts as shown in \autoref{tab:facets}. Weighted sums of linguistic features were used to measure these facets, where the coefficients were calculated in~\cite{crossley2019automated} for the corresponding correlations.

\begin{table}[h]
    \centering
\resizebox{\linewidth}{!}{
\begin{tabular}{@{}llll@{}}
\toprule
                  & Krippendorff's $\alpha$ & Cohen's kappa & Spearman's $\rho$ \\ \midrule
Understandability & 0.762                                 & 0.331                            & 0.420                                             \\
Consistency       & 0.434                                 & 0.315                            & 0.357                                            \\
Fluency           & 0.416                                 & 0.377                            & 0.345                                            \\
Diversity         & 0.681                                 & 0.458                            & 0.563                                            \\
Conciseness       & 0.641                                 & 0.482                            & 0.555                                            \\ \midrule
Perceived quality & 0.687                                 & 0.574                            & 0.483                                            \\ \bottomrule
\end{tabular}}
 \vspace{2mm}
    \caption{\textbf{Correlation between evaluation metrics and human raters.} Krippendorff's $\alpha$ and Cohen's kappa were used to measure the inter-rater reliability between two human raters on the formal quality and two other human raters on the perceived quality of 21 academic abstracts in computer science. Spearman's $\rho$ was used to measure the correlation between human raters and evaluation metrics. The perceived quality given by the evaluation metrics is the mean of its calculated formal quality.}
    \vspace{-3mm}
    \label{tab:correlation2}
\end{table}

\par \changed{To verify the validity of the evaluation metrics in \autoref{tab:facets}, we randomly selected $21$ academic abstracts in the \textit{CSAbstract} dataset~\cite{Cohan2019EMNLP} and retrieved their original articles. We defined the formal quality of the abstracts as five aspects measured by linguistic features, and the perceived quality was scored by experts directly after reading the abstracts. Following the annotation guidelines in \autoref{Rated Abstracts in the User Study}, we recruited four senior Ph.D. candidates in computer science to rate the abstracts, two of whom assessed formal quality and two assessed perceived quality. To assess the reliability of the ratings, we used Krippendorff's $\alpha$ and Cohen's kappa. As shown in \autoref{tab:correlation2}, we obtained inter-rater reliability (IRR) in the interval of $(0.4,0.8)$ and Cohen's kappa in the interval of $(0.3, 0.6)$, which indicates a moderate agreement among human raters. In addition, we assessed the correlation between the average human raters and the evaluation metrics using Spearman's $\rho$. The results showed moderate correlations between human and automatic metrics. Moderate correlations are acceptable because correlations between automatic evaluation metrics and human raters are usually not very high ~\cite{sai2022survey,wang2020asking}}.

\par In addition to experimentally verified correlations, correlations between linguistic features and corresponding quality are also meaningful in \autoref{tab:facets}. Usually, words that appear less frequently in the \textit{COCA spoken corpus}\footnote{\url{https://www.english-corpora.org/coca/}} or \textit{SUBTLEXus} corpus\footnote{\url{https://www.ugent.be/pp/experimentele-psychologie/en/research/documents/subtlexus}} are uncommon, so they can make the text less easily understood. Consistency refers to the factual alignment between
the abstract and the source and \textit{ROUGE-3} is illustrated to be related to consistency ~\cite{fabbri2021summeval}. By definition, \textit{ROUGE-3} is used to measure the source similarity, which can be interpreted as high source similarity implies a high consistency. Furthermore, the linguistic features listed in \autoref{tab:facets} correspond to \textit{Fluency} and are intuitively correlated. The implication of \textit{Diversity} is twofold: lexical diversity and syntactic diversity. The first eight features in the row are used to measure lexical diversity~\cite{kyle2021assessing}, and the other features are used to measure syntactic diversity~\cite{kyle2016measuring}. Finally, conciseness is also associated with the listed features, whose thresholds refer to the reference abstract.

\subsection{Design of User Interface}

\begin{figure*}[h]
 \centering % avoid the use of \begin{center}...\end{center} and use \centering instead (more compact)
   \vspace{-4mm}
 \includegraphics[width=\textwidth]{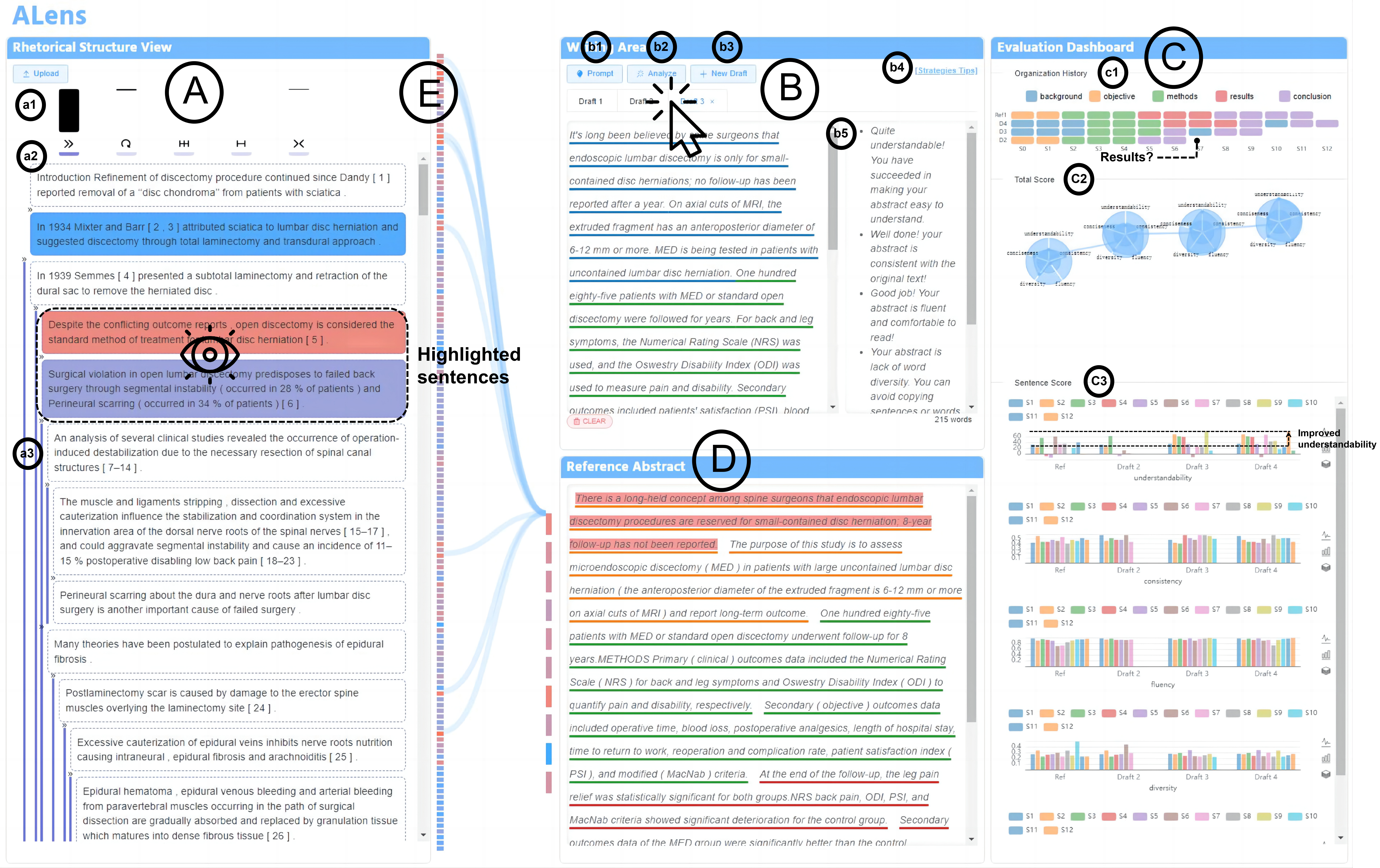}
 \vspace{-6mm}
 \caption{(A) The rhetorical structure view shows the RST tree of the article. (B) The writing area prepares the user to write an abstract. (C) The evaluation dashboard displays the results of semantic analysis results at different levels. (D)The reference abstract and (E) the flow map reveal the most relevant sentences from the reference abstract source.}
 \label{fig:overview}
   \vspace{-3mm}
\end{figure*}

\par Following the design principles mentioned in the formative study, we built \textit{ALens} as a responsive web-based application to demonstrate the academic abstract writing training process. The front-end interface includes a \textit{Rhetorical Structure View}, a \textit{Flow Map}, a \textit{Writing Area}, a \textit{Reference Abstract}, and an \textit{Evaluation Dashboard}. The \textit{rhetorical structure view} displays the parsed rhetorical tree of the original article and is designed to facilitate the user to quickly grasp the main information. After catching the main ideas, users can write their abstracts in the writing area. When users finish their drafts, they can utilize the NLP models to analyze their abstracts. The results of the analysis will be displayed on the \textit{evaluation dashboard}. Based on the evaluation results, users are expected to polish their abstracts. \textit{Reference Abstract} displays the results of the organization of the reference, and the most relevant sentences in the abstract are displayed through the \textit{flow map} and \textit{rhetorical structure view}.  

\begin{wraptable}{r}{5cm}
	\centering
	\begin{tabular}{lrrr}
\toprule
Rhetorical Relation & Glyph \\ \midrule
Background           & \begin{minipage}[b]{0.05\columnwidth}
                    		\centering
                    		\raisebox{-.5\height}{\includegraphics[width=0.5\linewidth]{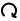}}
                    	\end{minipage}     \\
Contrast             & \begin{minipage}[b]{0.05\columnwidth}
                    		\centering
                    		\raisebox{-.5\height}{\includegraphics[width=0.5\linewidth]{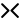}}
                    	\end{minipage}     \\
Elaboration          & \begin{minipage}[b]{0.05\columnwidth}
                    		\centering
                    		\raisebox{-.5\height}{\includegraphics[width=0.5\linewidth]{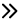}}
                    	\end{minipage}     \\
Joint                & \begin{minipage}[b]{0.05\columnwidth}
                    		\centering
                    		\raisebox{-.5\height}{\includegraphics[width=0.5\linewidth]{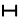}}
                    	\end{minipage}    \\
Sequence             & \begin{minipage}[b]{0.05\columnwidth}
                    		\centering
                    		\raisebox{-.5\height}{\includegraphics[width=0.5\linewidth]{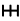}}
                    	\end{minipage}      \\ \bottomrule
	\end{tabular}
  \vspace{3mm}
\caption{Glyphs of rhetorical relations.}
 \vspace{-6mm}
\label{tab:rhetorical_glyph}
\end{wraptable}

\subsubsection{Rhetorical Structure View}

\par The Rhetorical Structure View (\autoref{fig:overview}A) is designed to help the user grasp the hierarchy of ideas and thus quickly identify the main ideas \changed{\textbf{(R2)}}. The rectangle (\autoref{fig:overview}A-a1) at the top of the view indicates the number of respective relations by their length. To make the rhetorical relations between sentences more intuitive, these relations are visually encoded with glyphs, as shown in \autoref{tab:rhetorical_glyph}. Learners can click on these glyphs (\autoref{fig:overview}A-a2) to hide the secondary sentences in a pair of relations. For example, by clicking on the elaboration glyphs, all elaborated sentences are preserved and the font color of the supporting sentences is lightened. In this way, learners can quickly capture key information, such as the core sentences in the elaboration relation and the secondary sentences in a contrast relation, or they can easily check the context by clicking on the glyphs again. In addition, we use a flattened tree structure (\autoref{fig:overview}A-a3) to display the hierarchy of ideas in the text, which is compact and makes good use of space. Sentences are wrapped as leaf nodes, logical glyphs are on the inner nodes, and the color depth of the rectangle (\autoref{fig:overview}A-a4) represents the number of corresponding relationships in the paragraph, so users can quickly identify the core sentences of each paragraph.

\par
\begin{figure}[h]
  \centering 
 % avoid the use of \begin{center}...\end{center} and use \centering instead (more compact)
 \includegraphics[width=0.48\textwidth]{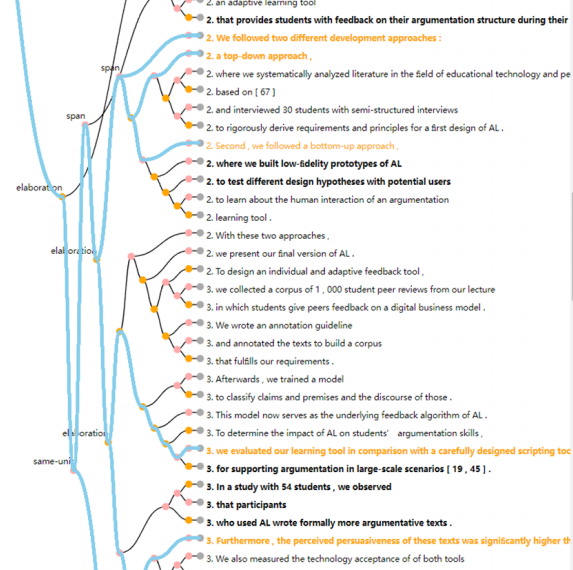}
 \vspace{-5mm}
 \caption{Design alternative of the rhetorical structure tree.}
\vspace{-4mm}
 \label{fig:alternative_rst}
\end{figure}

 \textbf{Design Alternative.} To represent the hierarchy of ideas, we initially designed the rhetorical structure tree, as shown in \autoref{fig:alternative_rst}. The leaf nodes are connected to sentences, and pop-up tooltips contain the names of the relations. However, during the design iteration, users commented that there was a large amount of white space on the left side of the tree and it was too cumbersome to move the mouse over the internal nodes to see the relations. In addition, they criticized that sentences in the article were placed side-by-side and the paragraph structure was broken, resulting in low readability. Therefore, we chose the current design to display the relationships visually and minimize the differences from the original natural text and ensure readability.

\subsubsection{Writing Area}
\par The writing area (\autoref{fig:overview}B) supports basic editing functionality, allowing users to write drafts manually or by clicking on the ``Prompt'' button (\autoref{fig:overview}B-b1), which uses an existing summarization model to predict drafts as prompts \textbf{(R5)}. Before writing, a suggested abstract writing strategy is provided by clicking on the ``Strategies Tips'' (\autoref{fig:overview}B-b4), enabling novices to quickly start abstract writing. After completing a draft, learners can click on the ``Analyze'' button (\autoref{fig:overview}B-b2) to analyze the abstract draft in six aspects (one for organizational structure and five for linguistic features). The sentences in the writing area are classified into five types and highlighted in different colors. Meanwhile, \autoref{fig:overview}B-b5 provides users with guided steps on how to improve the content and style of their abstracts based on the sentence classification results and the scores of linguistic features. Users can implement feedback by creating new drafts (\autoref{fig:overview}B-b3) to gradually improve the quality of their abstracts.

\subsubsection{Evaluation Dashboard}
\par The evaluation dashboard (\autoref{fig:overview}C) displays evaluation metrics at different granularities \textbf{(R1, R3)}. We design the organization map (\autoref{fig:overview}C-c1) as a row of aligned rectangle tiles -- the top row represents the organization scheme of the first draft, and the bottom row represents the most recent. Each tile in the row represents a sentence in the draft, and its color encodes the type of that sentence. After writing several drafts, users are expected to find the best organization scheme, and they are anticipated to identify the writing style of a group of papers by analyzing the best organization scheme for each paper in the group. In addition to the organization scheme, the line chart (\autoref{fig:overview}C-c2) records the overall score of the serialized drafts, with the five linguistic features encoded by the radar plot. However, the whole abstract and its scores for the five aspects may confuse learners~\cite{weigle2002assessing}, as they still need to recognize which parts need to be revised and which parts are already good. Therefore, (\autoref{fig:overview}C-c3) provides a more fine-grained analysis of the abstract. Each row represents a linguistic aspect, and a set of bars in the bar chart represents sentences in that draft. In this way, users can determine which sentences and aspects have not been considered and are poorly written. As a result, they can revise their drafts in a more precise and clear direction.

\subsubsection{Reference Abstract with a Flow Map}
\par The reference abstract with a flow map (\autoref{fig:overview}D\&E) is designed to reveal the writing style of the reference abstracts \textbf{(R4)}. Organizational detection is applied to the reference abstracts to explicitly reveal their organizational scheme. Also, we use a flow map to find the most relevant sentences in the source text for each sentence in the reference abstract. We use the sentence transformer~\cite{reimers-2020-multilingual-sentence-bert} to calculate the semantic similarity score of each sentence in the source text with each sentence in the reference abstract. Each square tiles on one side represent a sentence from the source text or the reference. The color depth of each tile on the abstract side is calculated by averaging the first $k$ similarity scores, where $k$ can be specified by the user. And the tiles on the source text side represent the similarity score when the user's mouse is placed over a sentence in the abstract. Meanwhile, the top $k$ similar sentences in the source text are highlighted and linked to the sentences in the reference. In this way, users can explore the writing patterns of reference abstracts. For example, they may find that the sentence in the reference abstract describing the background may come from the end of the paragraph introducing the background in the source text. In this way, knowledge about the writing style of the abstract can be constructed.

\section{Evaluation}

\par We evaluate the effectiveness of \textit{ALens} in two ways. First, we describe two usage scenarios with two target users of \textit{ALens}. Before that, we conducted a 10-minute tutorial with the involved participants to introduce \textit{ALens}. We then asked them to explore with \textit{ALens} for half an hour in a think-aloud manner. Second, we invited $21$ participants who had no exposure to our system to conduct a user study to further assess the potency of \textit{ALens}.

\subsection{Usage Scenario I}

\par In this subsection, we describe how Anker, a third-year undergraduate student, used \textit{ALens} to train his academic abstract writing skills. Anker is from the Department of Biomedical Engineering and has been starting his research career for about four months. Prior to using \textit{ALens}, he had no experience writing academic abstracts for journals and conferences, but he did have experience writing abstracts for course essays. We chose a paper from biomedical science and trained him in writing academic abstracts in a related field.

\begin{figure}[h]
 \centering
  \vspace{-3mm}
\includegraphics[width=0.5\textwidth]{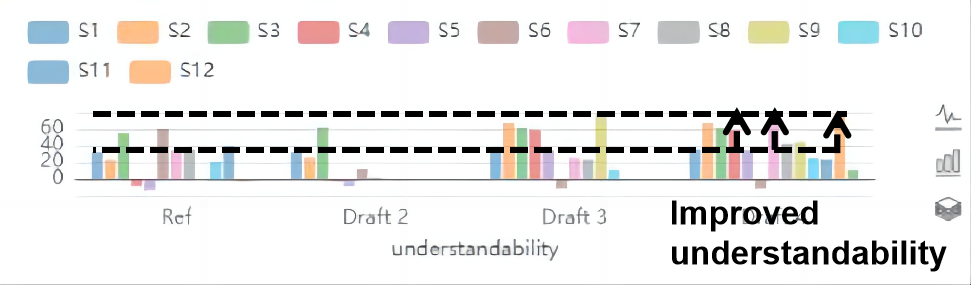}(a)
\includegraphics[width=0.5\textwidth]{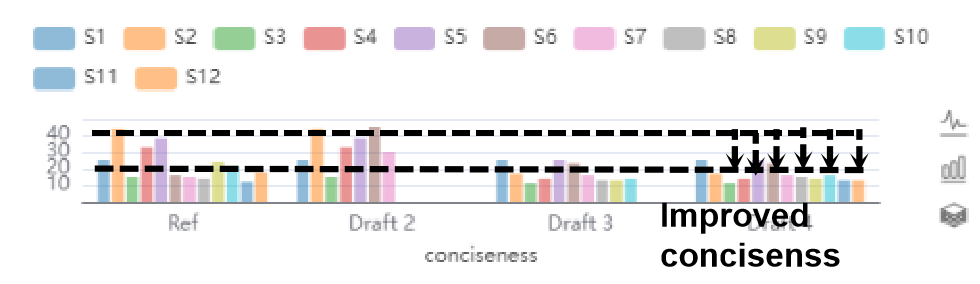}(b) 
\includegraphics[width=0.5\textwidth]{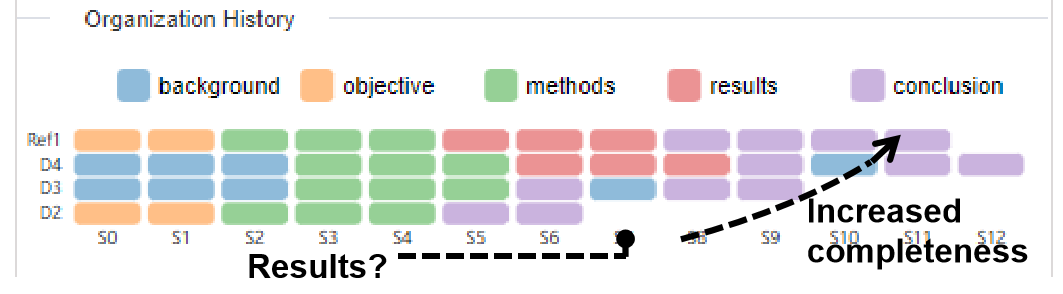}(c)
\vspace{-6mm}
 \caption{Anker's training procedure: (a) The results of comprehensibility and references for each sentence in the three analysis drafts are on the far left. (b) The conciseness of each sentence in the three analyzed drafts and the results of the analysis of the reference are on the leftmost side. Relatively short bars imply relatively concise sentences. (c) History of the content organization of the three analyzed drafts and the organization of the reference.}
 \label{fig:subviews}
 \vspace{-3mm}
\end{figure}

\begin{figure*}[h]
 \centering % avoid the use of \begin{center}...\end{center} and use \centering instead (more compact)
     \vspace{-3mm}
 \includegraphics[width=\textwidth]{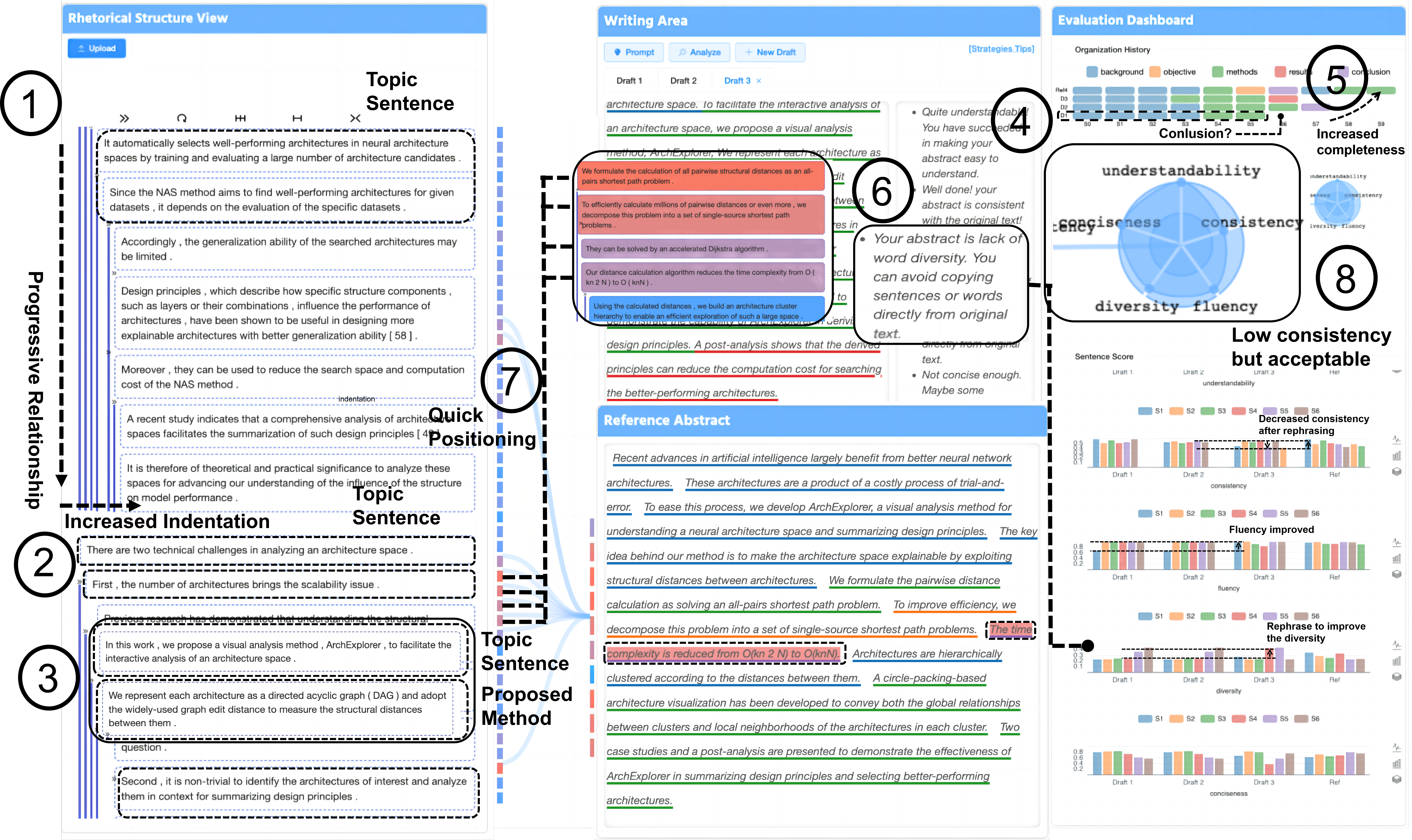}
 \vspace{-6mm}
 \caption{(1) The indentation before the sentence implies a progressive relationship in each paragraph. (2) Topic sentence and two juxtaposed challenges in the second paragraph. (3) The conclusion was omitted in the first draft. (4) The omission of the conclusion in the first draft was found, and content related to the assessment was added to improve the completeness of the abstract. (5) Based on the hint from \textit{ALens}, Jimmy rewrote the abstract and diversity was improved. (6) Descriptions of efficiency were quickly highlighted and positioned. (7) Jimmy confirmed that the low consistency of the abstract could be caused by the rephrasing of the reference abstract, so he considered that low consistency was acceptable.}
 \label{fig:case2}
  \vspace{-3mm}
\end{figure*}

\par First, he uploaded a prepared text file and then parsed the article into a rhetorical tree, as shown in~\autoref{fig:overview}(A). We observed that he started to consciously select sentences while reading the article, explaining \textit{``as far as I know, a typical abstract should present the purpose of the work, what problems it tries to solve, the research methods used, and the conclusions''}, and then he created a new tab to parse the copied sentences and clicked on the ``Analyze'' button (\autoref{fig:overview}(B)(b2)). Looking at the conciseness (\autoref{fig:subviews}(b)) and understandability (\autoref{fig:subviews}(a)) indicators (\autoref{fig:subviews}) in the evaluation dashboard (\autoref{fig:overview}(C)(c3), he found that his second draft was not easy to understand and not concise enough, mainly because of the problems with the last three sentences. In addition, he found that the last three sentences were too long to read, so he broke them into shorter sentences, added some connectives, and then clicked the ``Analyze'' button. The result of the analysis showed that the sentences in his draft were more concise and easier to read. Then he decided to look at the reference abstract. To his surprise, he found that he had missed the results of the experiment in the \textit{organization history view} (\autoref{fig:subviews}(c)). He said that \textit{``this is the first time I know that the experimental results needed to be independent of the conclusions, which I had previously thought already contained the results.''} Anker concluded with a quick review of the results section of the paper, grasping the comprehensive description of the results from the lengthy description of the results under the auxiliary rhetorical relations.

\subsection{Usage Scenario II}

\par \changed{In this subsection, we describe how Jimmy, a first-year graduate student, used \textit{ALens} to train his academic abstract writing skills. Jimmy comes from the Department of Computer Science and Engineering, and he has been starting his graduate career for about four months. Prior to using \textit{ALens}, he had no experience writing academic abstracts for journals or conferences, but he did have experience writing abstracts for course papers. We selected a paper from IEEE TVCG and trained him in writing academic abstracts in a related field. First, as shown in \autoref{fig:case2}, he uploaded the introduction section of the prepared article, and the system parsed it into a rhetorical tree. He first looked at each paragraph of the rhetorical tree, and based on the visualized structural information, he found that overall, the indentation of sentences in each paragraph was increasing (\autoref{fig:case2}(1)), which meant that each paragraph was largely in a progressive relationship, so he inferred that the first sentence of each paragraph was the main idea sentence, and the sentences that were indented too much were most likely not the abstract alternatives that needed to be focused on (\autoref{fig:case2}(2)). Next, he looked at the first paragraph, which was mainly about the general background of the article, and he found that the paragraph mainly revolved around the first two sentences (\textbf{Topic Sentence in \autoref{fig:case2}}), so he copied the first two sentences directly into the abstract. He continued with the second paragraph, which focused on two ``technical challenges'' in the related area (\textbf{Topic Sentence in \autoref{fig:case2}}). From the sentences with the same level of indentation, he intuitively found these two juxtaposed challenges, which he considered to be more important, and therefore copied them. In addition, he thought that transitions were also important logical relationships, but when he looked at the transitions and found the content after ``however'', he thought that it did not need to be included in the abstract because the content of the transition had already been mentioned in the first challenge. Moreover, when he wrote the first draft, he thought that the transitions all fell under the above two technical challenges. He turned to move on to the third paragraph, which focuses on the method proposed by the authors (\autoref{fig:case2}(3)), which he thought was the focus of the article and needed more space to discuss. By looking at the rhetorical tree, he found that the content of the third paragraph mainly revolved around the first sentence, so he extracted the first sentence. He further looked at the rhetorical tree and found that part of the content is the author's proposal to use the DAG method (\autoref{fig:case2}(3)) to characterize the problem, so he also extracted the DAG-related description words into the abstract. He was eager to use the above-extracted content as the first version of the abstract and clicked the ``Analyze'' button to see the results.}

\par \changed{According to the analysis result, \textit{ALens} thought that the abstract of this version lacked conclusive content (\autoref{fig:case2}(4)), and also the first sentence did not have enough fluency. In response to the first problem, Jimmy revisited the rhetorical tree and found that he had overlooked the assessment-related content, so he added the relevant content and revised the first sentence. After analyzing again, he found that the completeness of this version of the abstract was much improved (\autoref{fig:case2}(5)), but the fluency of the first sentence was not significantly improved. He checked the hints (\autoref{fig:case2}(6)) and learned that he might have retained too much content from the original text, so he made a proper rephrase of the abstract and then re-analyzed it. At this point, he found that the diversity of this version had improved, but the consistency with the article had decreased. He further guessed that he might have modified the original text, so the moderate decrease in consistency between sentences was acceptable. Overall, Jimmy was satisfied with this version of the abstract and did not intend to continue revising it.}

\par \changed{He then clicked ``Show Reference Abstract'' to see the gap between his written abstract and the reference abstract for further learning, and the system displayed the relevant indicators. He found that the organization of the reference abstract was clearly different from the version he finally submitted. Specifically, he found that the reference abstract consisted mainly of one type of sentence, i.e., the objective, so he looked at the relevant sentences of the reference abstract based on the color indicators and found a description for efficiency (\autoref{fig:case2}(7)), i.e., ``\textit{ Our distance calculation algorithm reduces the time complexity from $O(kn^2N)$ to $O(knN)$.}'', ``\textit{which is indeed what I missed in my abstract,}'' said Jimmy. He clicked on that sentence, and the rhetorical tree on the left side of the system showed the original content most relevant to that sentence by highlighting it. ``\textit{I can locate the relevant sentence in the original text very quickly.} (\autoref{fig:case2}(7))'' He looked at the most relevant part of the original text and found that it was indented to the same degree as the DAG method proposed by the author, so he inferred that they were true of the same importance, and ``\textit{that was ignored by me.}''}

\par \changed{He finally observed the total score of the reference abstract and found that the reference abstract scored very high in all categories except consistency . Using the ``quick locate'' function (\autoref{fig:case2}(7)), he compared each sentence of the reference abstract with the original text and found that the reference abstract had made a lot of rephrasing to the original text, so the consistency with the original text was not very high. At the same time, he noticed that the first sentence of the reference abstract only differed from the key content of the original text by one word, so this sentence had the highest consistency. ``\textit{This confirms my inference in my writing,}'' that is, rephrasing causes a slight decrease in consistency (\autoref{fig:case2}(7)).}

\subsection{User Study}

\par \changed{We quantitatively conducted a laboratory experiment to evaluate the performance of \textit{ALens} and compared it to a baseline training system.}

\par
\begin{figure}[h]
%  \centering 
 % avoid the use of \begin{center}...\end{center} and use \centering instead (more compact)
 \centering
 \includegraphics[width=0.5\textwidth]{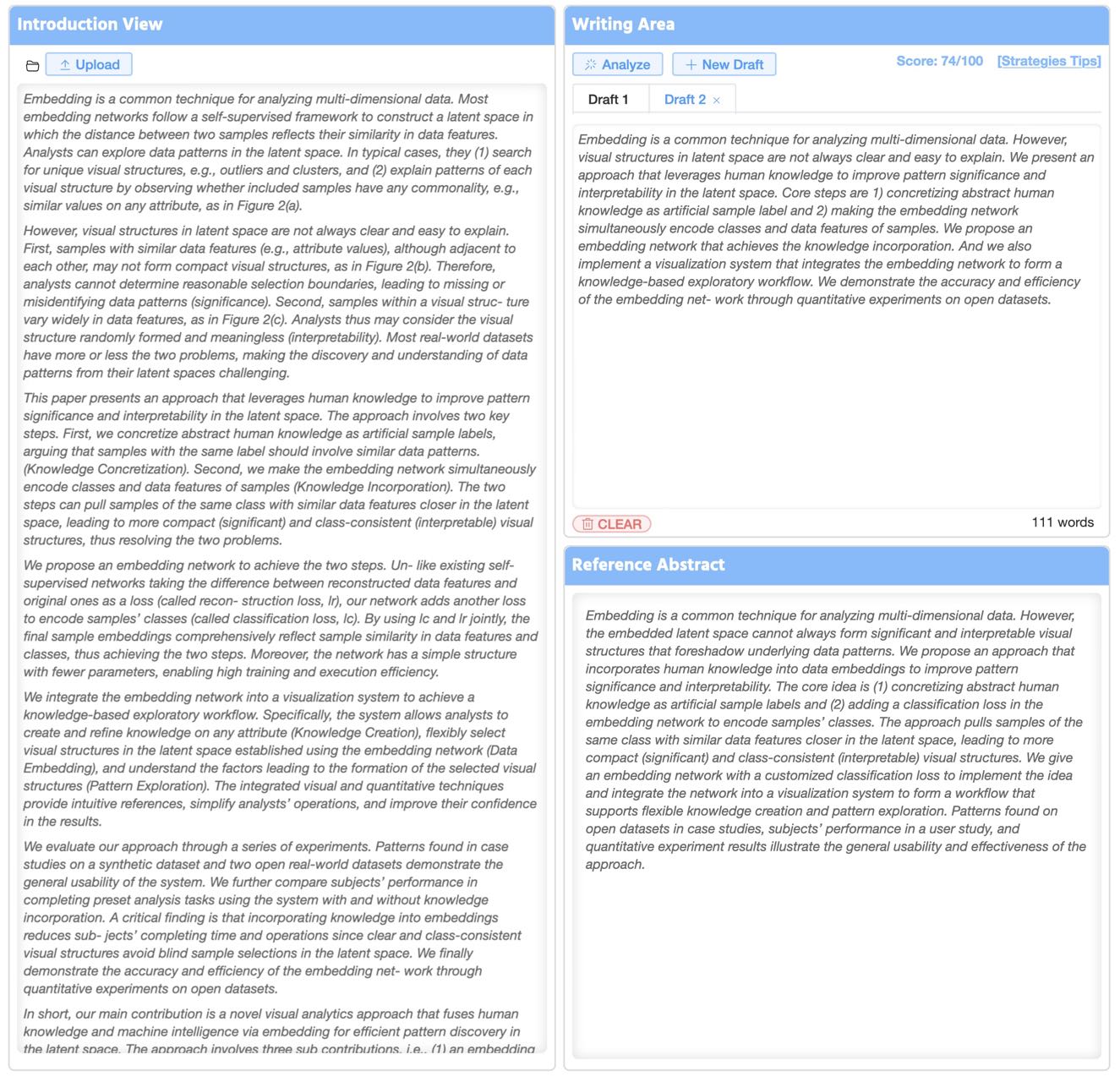}
  \vspace{-6mm}
 \caption{The baseline system supports users in reading articles (Introduction View), writing abstracts (Writing Area), and checking the reference abstract (Reference Abstract). The baseline can provide an overall score on the quality of the abstract.}
 \label{fig:baseline}
    \vspace{-3mm}
\end{figure}

 \textbf{Baseline Training System.} To evaluate \textit{ALens}, we built a baseline system, \changed{as shown in \autoref{fig:baseline}}, which simulates the traditional way of writing an abstract. We controlled for similarities and differences between \textit{ALens} and the baseline. Specifically, they both follow the same approach to abstract writing and share many features. First, both tools have a ``Strategies Tips'' button to learn general abstract writing strategies. In addition, a new draft button has been implemented to facilitate the iterative process of writing an abstract. The reference abstract appears when clicking on the ``Show Reference`` and disappears when clicking on the ``New Draft'' button. The difference between the baseline and \textit{ALens} is that in the baseline system, users reflect on their abstracts based on the same overall score and reference abstract, but in \textit{ALens}, they can obtain additional supporting information from the RST, classification, and metrics for revision.

\par \textbf{Hypotheses.} To answer \textbf{RQ4}, we investigated perceived usefulness and usability between participants who used \textit{ALens} and those who usedthe baseline system. In particular, we use a Wilcoxon rank sum tests to assess whether there are significant differences in the means of the constructs. Research on self-regulated learning theory suggests that personal feedback can help them learn better~\cite{bandura1997social}. In the learning process, self-reflection is important for effective learning, which can trigger the creation of new knowledge through self-regulated learning~\cite{zimmerman2001self}. Writing, as a creative process, is highly dependent on engagement~\cite{hirvela2013paraphrasing}. It has been argued that when students are attentive and engaged in the writing process, they are able to write in a more cohesive manner~\cite{hirvela2013paraphrasing}. Therefore, we propose the following hypothesis to answer \textbf{RQ5}. \textit{\textbf{H1}: Individual feedback helps users to write abstracts in a more appropriate style than the baseline system.} Here, style is defined by content organization and language style. \textit{\textbf{H2}: Compared to the baseline system, \textit{ALens} helps users to construct knowledge for academic abstract writing.} \textit{\textbf{H3}: Compared to the baseline system, \textit{ALens} improves user satisfaction with the final draft of the abstract.} \textit{\textbf{H4}: Compared to the baseline system, \textit{ALens} enables users to be more involved in the writing process.}

\par \textbf{Experiment Setup.} To test our hypotheses, we designed a laboratory experiment in which participants were asked to read and comprehend the introduction of a given article in the field of computer science, write an abstract based on the introduction, learn the writing style of the given reference abstract, and \changed{revise the abstract they wrote at least once.} Since academic abstract writing is highly relevant to the field, and students outside the field usually encounter obstacles in reading and understanding articles, we recruited $21$ students from the Department of Computer Science of a local university via social media. Participants were randomly assigned to an experimental and control group. The experimental group used \textit{ALens}, while participants in the control group used the baseline system. After random assignment, there are $12$ students in the experimental group and $9$ students in the control group. Participants in the experimental group ($9$ males and $3$ females) had an average age of $21.17$ ($SD=1.64$) and they had an average of $0.33$ ($SD=0.65$) of academic abstract writing experience. In the control group, there were $7$ males and $2$ females. Their mean age was $20.89$ ($SD=1.54$) and they had written an average of $0.22$ ($SD=0.67$) academic abstracts. Upon completion, each participant received a $\$20$ stipend for their contribution. We designed an experiment with three phases, namely, a \textbf{pre-test phase}, a \textbf{short-term abstract writing training phase}, and a \textbf{post-test phase}. During the training phase, the experimental group used \textit{ALens} and the control group used the alternative baseline system for reading, writing, and learning.

\par \textbf{Pre-test Phase.} To test whether our initial random grouping was indeed random, we first tested participants' acceptance of new information technology, feedback seeking, self-confidence, and academic abstract writing skills through a 16-question pretest. First, \changed{we asked participants four questions about the acceptance of new information technology for writing assistance, referring to the approach proposed by Agarwl et al.~\cite{agarwal2000time}.} \changed{Second, we based on Ashford et al.~\cite{ashford1986feedback} and asked them questions about the ability to actively seek feedback on academic paper writing.} In addition, based on Ashford et al.~\cite{ashford1986feedback}, \changed{we tested their ability to control their mental ability states with the aim of knowing whether they were overconfident in reporting their experimental results. Samples items for the constructs are \textit{``I don't believe in myself''; ``I feel that I am a valuable person on an equal footing with others''; ``I seem to have a real inner strength when dealing with things. I have a very solid foundation, which makes me very confident in myself.''}} Fourth, \changed{studies~\cite{johnstone2002effects,tuan2010enhancing,hasani2017using} have shown that juniors with task-specific writing practice (e.g., academic writing) statistically performed better in writing than students with only general writing training. Therefore, we indirectly understand their ability to write academic abstracts by asking them about their experience in writing academic abstracts such as past academic writing achievements and problems pointed out by reviewers or instructors during research submissions.}

\par \textbf{Short-Term Abstract Writing Training Phase.} Before this phase began, we gave a brief introduction to our system and let them play with it for about $5$ minutes. To test whether our system enables users to construct knowledge of academic abstract writing, we developed a short-term abstract writing training phase as follows. First, participants were asked to read the introduction of a computer science paper from the IEEE Transactions on Visualization and Computer Graphics (TVCG), since the introduction of the TVCG paper usually contains all the information of the article. Our co-author, an expert in the field of visualization, confirmed this fact. Participants were asked to spend at least $5$ minutes reading the introduction of about $800$ words to ensure that they had a basic understanding of the article. They were then asked to spend at least $10$ minutes writing the abstract. Subsequently, users were allowed to check the reference abstract and learn the writing style of that abstract for a minimum of $5$ minutes. The experimental group was then asked to use \textit{ALens} to check the organization of the abstract and the placement of the core sentences, while students in the control group were allowed to analyze the abstract based on their knowledge. Then, if they were in the experimental group, they needed to revise their first drafts based on what they had learned about writing from the reference abstract and based on the evaluation metrics. Finally, they could progressively embellish the abstract until they were satisfied with the draft. All drafts from the training process were collected by both systems and sent for post-evaluation.

\begin{table*}[h]
    \centering
\resizebox{\linewidth}{!}{
\begin{tabular}{@{}lllllllll@{}}
\toprule
                                      & Content Integrity & Content Organization & Comprehensibility & Consistency & Fluency & Diversity & Conciseness & Perceived quality \\ \midrule
Krippendorff's $\alpha$ between two raters & 0.953             & 0.903                & 0.682             & 0.834       & 0.732   & 0.694     & 0.754       & 0.676             \\
Cohen's kappa between two raters      & 0.807             & 0.851                & 0.637             & 0.702       & 0.643   & 0.578     & 0.602       & 0.631             \\ \bottomrule
\end{tabular}}
\vspace{1mm}
    \caption{\changed{\textbf{IRR between human raters.} Two human raters for formal quality and two other human raters for perceived quality.}}
    \label{tab:correlation1}
    \vspace{-6mm}
\end{table*}

\begin{table*}[h]
\resizebox{\textwidth}{40mm}{
\begin{tabular}{@{}cccccc@{}}
\toprule
\multicolumn{6}{c}{\textbf{Pre-test Phase}} \\ \midrule
Group         & \begin{tabular}[c]{@{}c@{}}New technology \\ acceptance\end{tabular} & \begin{tabular}[c]{@{}c@{}}Self-\\ confidence\end{tabular} & \begin{tabular}[c]{@{}c@{}}Feedback\\  seeking\end{tabular} & \multicolumn{2}{c}{\begin{tabular}[c]{@{}c@{}}Times for\\  writing abstract\end{tabular}} \\
Mean ALens    & 5.83                                                                 & 5.42                                                       & 5.25                                                        & \multicolumn{2}{c}{0.33}                                                                  \\
Mean Baseline & 5.89                                                                 & 5.33                                                       & 5.00                                                        & \multicolumn{2}{c}{0.22}                                                                  \\
SD ALens      & 0.58                                                                 & 0.90                                                       & 0.87                                                        & \multicolumn{2}{c}{0.65}                                                                  \\
SD Baseline   & 0.60                                                                 & 0.87                                                       & 1.00                                                        & \multicolumn{2}{c}{0.67}                                                                  \\

Asymp.Sig. (2-sided)      & 0.831                                                                & 0.874                                                      & 0.433                                                      & \multicolumn{2}{c}{0.500}                                                                 \\ \midrule
\multicolumn{6}{c}{\textbf{Training Phase}}                                                                                                                                                                                                                                                                          \\\toprule
Group         & First Draft Formal quality                                           & First Draft Perceived quality                              & Second Draft Formal quality                                 & \multicolumn{2}{c}{Second Draft Perceived quality}                                        \\
Mean ALens    & 4.08                                                                 & 4.50                                                       & 5.39                                                        & \multicolumn{2}{c}{5.58}                                                                  \\
Mean Baseline & 4.22                                                                 & 4.44                                                       & 4.56                                                        & \multicolumn{2}{c}{4.78}                                                                  \\
SD ALens      & 0.74                                                                 & 0.90                                                       & 0.99                                                        & \multicolumn{2}{c}{0.79}                                                                  \\
SD Baseline   & 1.15                                                                 & 0.73                                                       & 0.74                                                        & \multicolumn{2}{c}{0.67}                                                                  \\

Asymp.Sig. (2-sided)       & 0.776                                                                & 0.874                                                     & 0.041                                                       & \multicolumn{2}{c}{0.025}                                                                 \\\toprule
\multicolumn{6}{c}{\textbf{Post-test Phase}}                                                                                                                                                                                                                                                                         \\\toprule
Group                                                        & First Draft Satisfaction                                   & Second Draft Satisfaction                                  & Engagement                             & Knowledge Construction                            \\
Mean ALens                                                                  & 3.50                                                       & 5.33                                                        & 4.33                                  & 5.33                                              \\
Mean Baseline                                                                 & 3.67                                                       & 4.67                                                        & 4.44                                  & 3.78                                              \\
SD ALens                                                                       & 1.09                                                       & 0.49                                                        & 0.49                                  & 0.98                                              \\
SD Baseline                                                                  & 0.71                                                       & 0.71                                                        & 0.53                                  & 0.67                                              \\
Asymp.Sig. (2-sided)                                                                 & 0.625                                                     & 0.026                                                       & 0.613                                & 0.001                                             \\ \bottomrule
\end{tabular}}
\vspace{1mm}
\caption{Results of statistical analyses of the \textit{ALens} and baseline systems on the Likert scale (1: low, 7: high).}
\label{tab:statistic_analysis}
\vspace{-8mm}
\end{table*}

\par \textbf{Post-test Phase.} In this phase, we first measured users' intention to use our system, as well as usability and usefulness after technology acceptance testing~\cite{vom2018future}. In addition, we measured user satisfaction with their first and final drafts, and measured perceived engagement. The sample items for the five constructs are \textit{``I will use the system for abstract writing training if it were released;'' ``I can write abstracts in the appropriate style;'' "I feel I can learn to use the system quickly;'' ``I am satisfied with the first draft I wrote using the system;'' ``I focused on the writing itself and the time passes quickly for me.''} We used a 7-point Likert scale (7: very sure, 1: not very sure, 4 for neutral statements) for participants to assess.

\par \textbf{Measurement.} Technology acceptance, user satisfaction, and engagement were used to evaluate the system from the user's perspective and to test hypotheses \textbf{H3} and \textbf{H4}. In addition, we tested hypotheses \textbf{H1} and \textbf{H2} by measuring the quality of abstracts from the two groups. We measured the quality of the drafts in two ways: 1) perceived quality and 2) formal quality. In particular, for perceived quality, we invited two senior researchers in the field of visualization to help us evaluate the abstracts on a Likert scale of 1 -- 7 (7: very good, 1: very poor). Both of them have $5$ years of research experience. We used their average score as the final score of the draft. For another, we analyzed the formal quality of the first and final drafts. We defined the formal quality of the abstracts as the following seven aspects: \textit{content integrity}, \textit{content organization}, \textit{comprehensibility}, \textit{consistency}, \textit{fluency}, \textit{diversity}, and \textit{conciseness}. \changed{The first two metrics are used to assess the classification results in terms of content integrity and content organization. The other five metrics are the same as the five metrics previously mentioned in \autoref{tab:facets}.} We used a Likert scale of 1 -- 7 (7: very good, 1:very poor) to create criteria for these seven aspects. The rating guideline can be referred to \changed{\nameref{Rating Guideline} in \autoref{Rated Abstracts in the User Study}.} We then annotated the $42$ ($21*2$) drafts ourselves and used our average score as the final score for the draft in that area. \changed{Krippendorff's $\alpha$ in \autoref{tab:correlation1} shows the resulting inter-rater agreement reliability (IRR) scores. We obtained Krippendorff's $\alpha$ scores between (0.67, 0.96) for the seven metrics, indicating considerable agreement between the two raters. In addition, Cohen's kappa was between (0.57, 0.71), showing the same result. Therefore, we conclude that the rated abstracts for the seven metrics should be reliable.}

\par To answer research questions RQ4 -- RQ5, we first ensured that our random assignment was successful and controlled for potential effects of small samples, and we compared the differences between the two groups on the four aspects in the pre-test phase as shown in \autoref{tab:statistic_analysis}. \changed{The two-sided asymptotic significance was greater than $0.05$ for all four constructs, which ensured that the two groups did not differ significantly on these four constructs. Since the distribution of the two groups was not always normal, we performed post hoc Wilcoxon rank sum tests for pairwise group comparisons.}

\subsubsection{\textbf{RQ4: What is the technology acceptance level among junior researchers?}}

\par
\begin{figure}[h]
%  \centering 
 % avoid the use of \begin{center}...\end{center} and use \centering instead (more compact)
 \centering
 \includegraphics[width=0.4\textwidth]{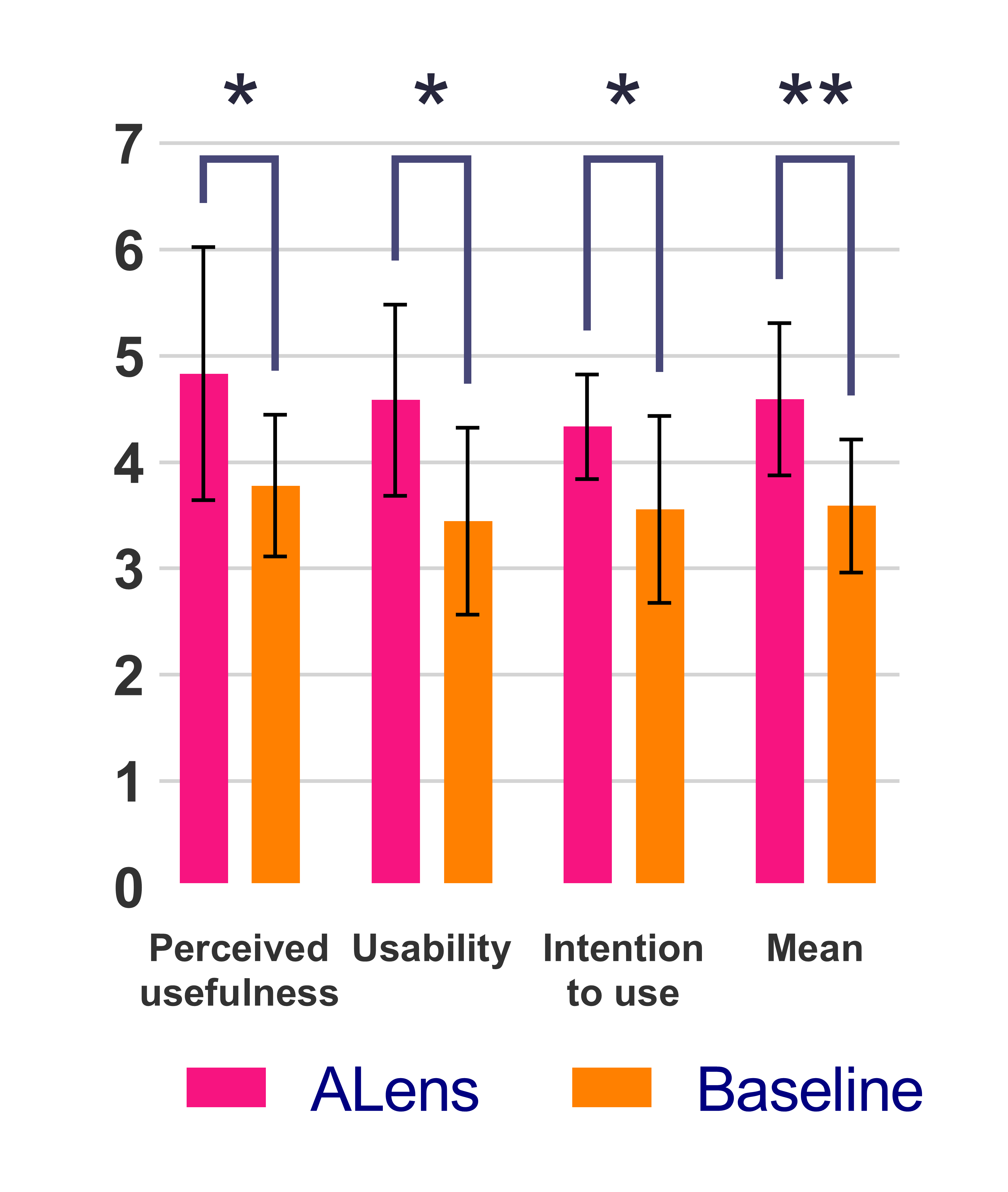}
 \vspace{-6mm}
 \caption{The technology acceptance of ALens and the baseline system on the Likert scale (1: low, 7: high). ($*: Asymp.Sig.<.05; **: Asymp.Sig.<.01$)}
  \vspace{-3mm}
 \label{fig:acceptance}
\end{figure}

 \autoref{fig:acceptance} shows the average user ratings for technology-related questions.  Wilcoxon rank sum tests show significant differences  in perceived usefulness ($Asymp.Sig. = 0.015$), usability ($Asymp.Sig. = 0.016$), and intention to use ($Asymp.Sig. = 0.039$) when abstract writing skills were trained with different systems. We averaged the Likert scores for perceived usefulness, intention to use, and usability to compare technology acceptance. We find that \textit{ALens} obtained a higher acceptance than the baseline system. Technology acceptance of a learning tool is an essential basis for further user learning. A positive technology acceptance provides a promising result for using this tool as an adaptive feedback application.

\subsection{Results}

\subsubsection{\textbf{RQ5: How effective is ALens in helping users write abstracts compared to the baseline system?}}
\ 
\par \textbf{Writing in a More Appropriate Style.} To test whether users write abstracts in a more appropriate style, we collected $42$ draft abstracts written by participants and assessed their formal and perceived quality. From \autoref{tab:statistic_analysis}, we can see that there was no significance between the first drafts of the two groups \changed{on formal quality ($Asymp.Sig. = 0.776$) and perceived quality ($Asymp.Sig. = 0.874$). However, there is a statistically significant difference (formal quality: $Asymp.Sig. = 0.041$; perceived quality: $Asymp.Sig. = 0.025$) between the second draft of the two groups, which demonstrates the effectiveness of \textit{ALens} in training users to write abstracts in a more appropriate style.} Therefore, \textbf{H1 is accepted}.

\par \textbf{Abstract Writing Knowledge Construction.} To test whether the users gained knowledge about abstract writing, we asked them three quantitative problems and one qualitative problem. \changed{The three quantitative questions were: 1) \textit{``I have a general understanding of what an abstract should include when using the system}''; 2) \textit{``I have a general understanding of how an abstract should be organized when using the system}''; and 3) ``\textit{I am now more familiar with the language style of the abstract.}''  The qualitative question was ``\textit{do you have new insights into abstracts? Can you talk about them?}`` The results of the three quantitative questions ($Asymp.Sig. = 0.001$) (\autoref{tab:statistic_analysis}) and the positive feedback from the qualitative question indicate a promising result that users can gain knowledge about abstract writing. Sample answers to the qualitative questions were \textit{``background is important and is usually described in one sentence}''; ``\textit{it is important to balance the proportion of the padding information}''; ``\textit{I first realize that there is a trade-off between diversity and consistency. While the phrases in the abstract should not be exactly the same as in the manuscript, preserving the original expressions is not a bad thing, given the consistency}''}. Therefore, \textbf{H2 is accepted}.

\par \textbf{Increased Satisfaction Level.} The results of the post-test phase showed that \textit{ALens} and the alternative tool enabled users to improve their satisfaction with their drafts (\autoref{tab:statistic_analysis} post-test phase). Although iteration usually makes things better, the difference in satisfaction ($Asymp.Sig. = 0.026$) between the two groups became more significant in the second draft, implying that \textit{ALens} could increase their satisfaction significantly. \textbf{H3 is accepted}.

\par \textbf{Higher Involvement in the Writing Process.} The statistical results showed that there was no significance ($Asymp.Sig. = 0.613$) between the two groups in terms of engagement. Therefore, \textbf{H4 is rejected}. We believe that users feel distracted during the long reading and writing process, and the interaction time for learning writing style and analyzing their drafts is relatively shorter.  

\par \textbf{Qualitative Feedback.} We also included some open-ended questions in the survey to get some suggestions for improvement. For example, we asked, \textit{which part(s) of the system need(s) improvement and why?} In general, most participants were positive about \textit{ALens}, especially the flow map, sentence-level evaluation, `` Strategies Tips``,  the sentence classification function, and the ''Prompt'' function. However, some participants also made constructive suggestions. In general, they complained about sometimes misleading rhetorical trees, incorrect sentence classification results, and confusing scores on the evaluation dashboard.

\section{Discussion and Limitation}
\subsection{Validity and Technology Acceptance Evaluation}
\par The outcomes of our user study demonstrated that the provision of adaptive formative feedback on students' abstract drafts significantly contributes to their ability to compose abstracts in a more appropriate style, encompassing content organization and language usage. This improvement was verified through assessments of both formal and perceived quality, where the final drafts of both student groups surpassed the initial drafts in terms of overall quality. We posit that this effect can be explained by the principles of self-regulated learning theory. However, upon comparing the quality levels between drafts from the same batch, we observed a notable increase in the discrepancy between the two student groups. This underscores the significance of delivering feedback in the correct proportion and granularity, as learner uptake and self-regulation are heavily influenced by these factors, as discussed by Bandura~\cite{bandura1997social}. In contrast to the alternative tool, \textit{ALens} offers a range of personalized feedback, thereby motivating students to modify their writing behaviors. The short-term enhancements witnessed in the user study regarding academic abstract writing provide compelling evidence that self-regulation fosters participants' motivation to acquire writing skills and construct pertinent knowledge. Furthermore, to effectively implement our study in a real-world scenario for abstract writing training, we conducted a validation of system technology acceptance, yielding promising results.

\subsection{LLM Impact on L2 Students' Academic Abstract Writing}
\par The recent development of Large Language Models (LLMs) has had a positive impact on L2 students' ability to learn academic abstract writing. For instance, these models can be used to automatically generate concise and accurate abstracts, helping students quickly grasp the core points and conclusions of their papers. They can provide targeted writing guidance, assisting students in organizing the structure and content framework of their papers. Additionally, these models can identify and correct language errors and grammar issues in students' writing, providing real-time feedback and suggestions to improve the language quality of their papers. By analyzing the abstracts generated by these models, students can learn excellent writing styles and structures, enhancing their academic expression abilities.
\par However, it's important to note that the improvement in these learning abilities is independent of the models themselves and requires students to invest additional time in comparison and comprehension. Otherwise, if students overly rely on the models to complete their abstract writing tasks, they may lose their ability to think independently and solve problems~\cite{yoon2004esl,deng2018deep}. Academic writing requires deep thinking and independent research, and excessive dependence on large language models may result in students lacking a profound understanding of the issues and the ability to think critically. Furthermore, the content generated by large models may not always be accurate and reasonable. Students need to possess critical thinking and judgment skills when using these models in order to correctly evaluate and apply the generated content. If students lack these skills, they may blindly accept the suggestions and guidance from the models, which can affect the quality of their writing and their academic expression abilities.

\subsection{Design Implications}
\par We conducted a formative study aimed at gaining insights into the challenges faced by L2 junior students/researchers during the process of writing academic abstracts. Based on our findings, we derived design requirements that are relevant to this context. To the best of our knowledge, \textit{ALens} represents one of the pioneering studies that have successfully established validated design requirements for an adaptive learning tool targeting academic abstract writing. Our research holds the potential to serve as a source of inspiration for individuals interested in the development of tools for training metacognitive skills. Instructors and developers of such tools can leverage our design requirements and discoveries to create their own training resources tailored specifically for enhancing academic abstract writing abilities.

\par The majority of existing computer-assisted writing tools primarily focus on assisting students in producing well-crafted writing pieces through iterative feedback and instructional support~\cite{strobl2019digital}. Similarly, in the present study, \textit{ALens} and other similar tools explicitly incorporate knowledge construction as a key objective during their design process. While these tools anticipate users to enhance their writing skills through computer-assisted guidance, informed by the principles of self-regulated learning theory~\cite{zimmerman2001self}, the specific factors contributing to user progress have not been adequately measured or fully elucidated within the theoretical framework. However, as an adaptive training tool, \textit{ALens} not only aids L2 junior researchers in achieving satisfactory writing outcomes but also endeavors to facilitate their exploration and acquisition of writing and stylistic knowledge. Notably, rather than providing a singular evaluation outcome such as scores or reviews, \textit{ALens} offers a carefully designed pipeline that enables users to delve into the underlying reasons behind the evaluation results. In addition to the usage scenarios delineated in this study for L2 junior researchers, we envision \textit{ALens} being of assistance to \textbf{experienced researchers} in analyzing abstracts outside their specialized domains. For instance, when sociologists or psychologists intend to contribute to the CHI community, \textit{ALens} can be a potential option for comparing abstract writing methodologies in both engineering and humanities, subsequently facilitating the transfer of writing knowledge across fields. In addition to promoting interdisciplinary learning, \textit{ALens} can be employed to acquire the abstract writing style employed by scientists or a specific research group of interest.

\subsection{Limitation}
\par Our study is subject to several limitations. First, there were concerns raised by four participants regarding the potentially misleading nature of the rhetorical tree employed in our research. Specifically, the performance of the tree structure in capturing relations such as ``Joint'' and ``Sequence'' was viewed as suboptimal. It is worth noting that our RST parser primarily operated at the sentence and paragraph levels. Due to the limited availability of sentence-level rhetorical structure datasets and the usage of the RST Discourse Treebank by Lynn and Marcu~\cite{lynnmarcu2002}, which is annotated at the EDU level, we merged the EDUs to obtain corresponding sentences. During the user study, it became apparent that the reason behind the participants' positive perception of the system's performance was primarily rooted in their modest expectations regarding full automation. As one participant stated, ``\textit{the tip itself is just a reference, after all, I still need to read through INTRODUCTION.}'' Second, five participants expressed dissatisfaction with the sentence classification model's effectiveness in correctly identifying categories such as ``objective'' and ``background'', as well as ``background'' and ``conclusion''. This issue can be attributed to the lack of high consistency between the annotated criteria and the original abstract. Moreover, the limited dataset available for abstract sentence classification restricts the generalizability of the model. Third, participants exhibited confusion regarding the assigned scores. We believe that a disparity exists between the human understanding of words and the descriptive metrics employed for evaluation purposes. Last, although our user study indicated that \textit{ALens} contributed to the improvement in abstract quality, it remains uncertain whether this enhancement can be solely attributed to increased editing. The study solely focused on written abstracts, with some observed editing behaviors, while the complete extent of editing actions was not documented or measured. In future investigations, we plan to record comprehensive edit logs throughout the entire process and analyze the effectiveness of editing behaviors.

\par Furthermore, during the user study, we successfully confirmed the immediate favorable impact of \textit{ALens} on participants' composition of academic abstracts. However, the long-term learning effects necessitate further validation. To address this, we plan to conduct a field experiment aimed at examining the efficacy and acceptance of \textit{ALens} in a practical setting. This experiment will involve the formation of two distinct groups: a control group that will receive feedback exclusively from a tutor, and an experimental group that will receive feedback from both the tutor and \textit{ALens}. By comparing the outcomes of these two groups, we aim to ascertain the long-term effectiveness of \textit{ALens} in facilitating and enhancing the process of academic abstract writing.

\section{Conclusion and Future Work}
\par This study introduces \textit{ALens}, an innovative automated feedback learning tool designed to enhance academic abstract writing training by integrating visualization and interactive elements. A comparative analysis between \textit{ALens} and a baseline system was conducted in a user study. The findings revealed that participants utilizing \textit{ALens} exhibited improved abstracts in terms of content organization and language style. Notably, \textit{ALens} demonstrated promising levels of technology acceptance and validity, thereby indicating its potential for practical application in academic abstract training. Furthermore, the outcomes of both the formative study and the user study offer valuable insights for informing the development of abstract writing training tools.

\par For further research, we present two prospective scenarios aimed at enhancing the efficacy of our RST model and sentence classification. Regarding RST, our intention is to adopt the approach outlined in~\cite{kobayashi2021improving} to refine our RST parser. This involves initially training the parser on automatically annotated data and subsequently fine-tuning it using the RST-DT corpus introduced by Carlson et al.~\cite{carlson2003building}. Additionally, we contemplate leveraging the capabilities of pre-trained language models, following the methodology proposed by Yu et al.~\cite{yu2022rst} in their work on RST. To improve the precision of sentence classification, we propose the utilization of a similar data augmentation technique. Specifically, we will commence by training our Bert model on automatically annotated data, and subsequently fine-tune it using a dataset that is specific to the relevant domain. In terms of the evaluation dashboard's confusing score, our objective is to enhance the comprehension of these scores by providing annotated exemplars and explanations. This approach aims to elucidate the underlying meaning of the scores and facilitate the development of actionable step-by-step guides for achieving higher scores. Furthermore, we are considering the exploration of a transformer-based model with the aim of predicting more accurate scores.

\begin{acks}
This work is partially supported by the Shanghai Frontiers Science Center of Human-centered Artificial Intelligence (ShangHAI) and Key Laboratory of Intelligent Perception and Human-Machine Collaboration (ShanghaiTech University), Ministry of Education.
\end{acks}

\bibliographystyle{ACM-Reference-Format}
\bibliography{sample-base}

\appendix
\section{Hyperparameters  of NLP models}
\label{appendixA}
\label{Hyperparameters  of NLP models}
\begin{table}[H]
    \centering
\resizebox{0.7\linewidth}{!}{
\begin{tabular}{@{}llllll@{}}
\toprule
Hyperparams                                                                              & PubMed & arXiv-cs \\  \midrule
Size of the intermediate feed forward layer in each T5Block & 2816   & 3072        \\
Size of the encoder layers                                  & 1024   & 768            \\
Maximum sequence length                                     & 16384  & 4096            \\
Number of attention heads                                   & 16     & 12            \\
Number of hidden layers                                     & 24     & 12          \\
Vocabulary size                                             & 32100  & 32100         \\
Size of the key, query, value                               & 64     & 64             \\
Number of decoder layers                                    & 24     & 12       \\
Dropout rate                                                & 0.1    & 0.1         \\
Activation function                                         & relu   & relu     \\ \bottomrule
\end{tabular}}
    \caption{Hyper-parameters of two summarization models based on \textit{LongT5} on the \textit{PubMed} and\textit{arXiv-cs} datasets.}
    \label{tab:hyperparams_summarization}
\end{table}

\begin{table}[H]
    \centering
\resizebox{0.7\linewidth}{!}{
\begin{tabular}{@{}lll@{}}
\toprule
Hyperparams                & PubMed 200k RCT & CSAbstract \\ \midrule
Number of attention heads  & 12              & 12         \\
Number of hidden layers    & 6               & 6          \\
Size of the encoder layers & 3072            & 3072       \\
Activation function        & glue            & glue       \\
Maximum sequence length    & 512             & 512        \\
Dropout rate               & 0.2             & 0.2        \\
Vocabulary size            & 30522           & 30522      \\ \bottomrule
\end{tabular}}
    \caption{Hyper-parameters of sentence classification model, \textit{BERT}, on  the \textit{PubMed 200k RCT} \textit{CSAbstract} dataset.}
    \label{tab:hyperparams_classification}
\end{table}

%\section{Correlation between Automatic Evaluation Metrics and Human Raters}

\section{Rating Guideline}
\label{Rated Abstracts in the User Study}
\label{appendixB1}
\label{Rating Guideline}

\begin{figure}
%  \centering 
 % avoid the use of \begin{center}...\end{center} and use \centering instead (more compact)
 \centering
 \includegraphics[width=\textwidth]{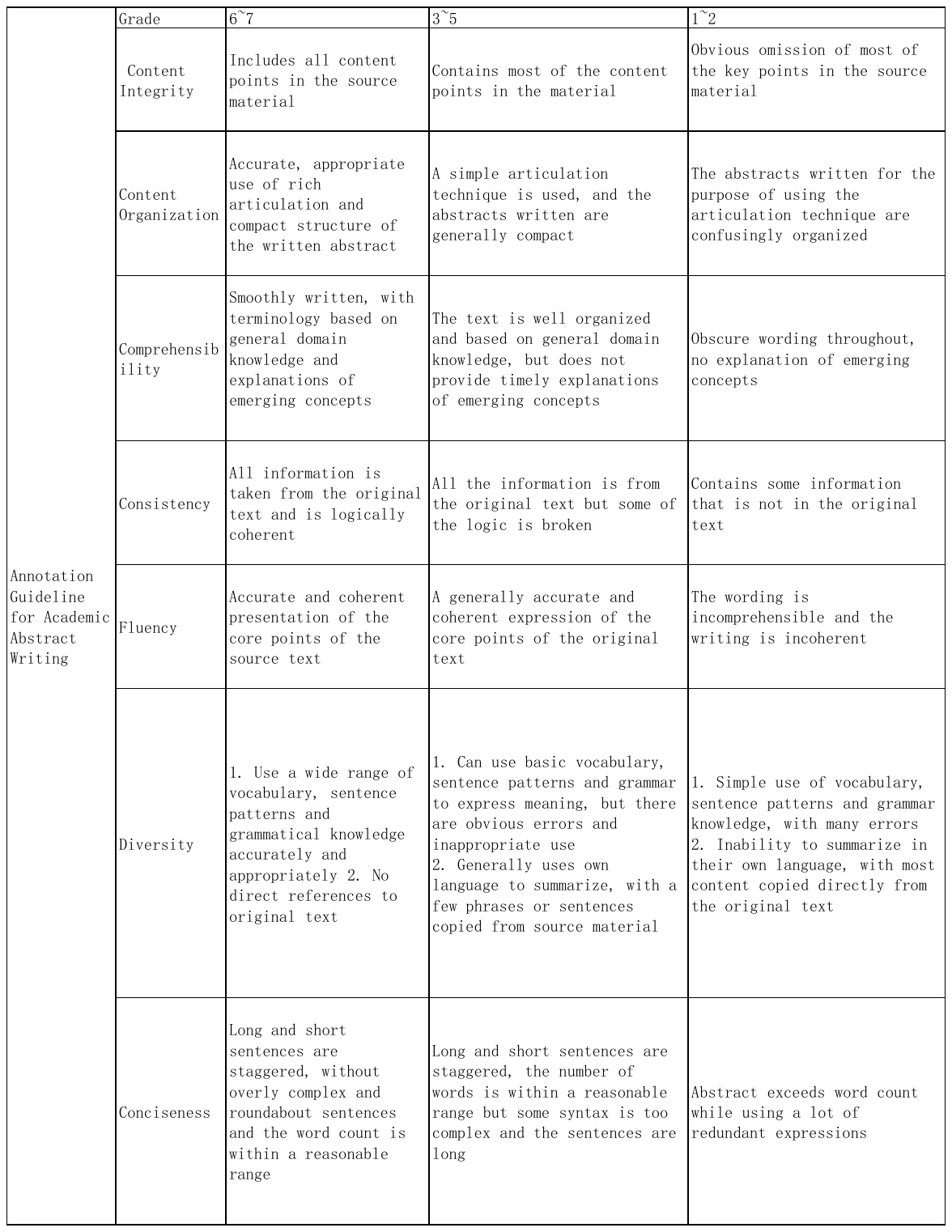}
  \vspace{-3mm}
 \caption{Guidelines for scoring formal quality in the user study. For formal quality, raters should rate abstracts according to these guidelines. For perceived quality, these guidelines are only a reference, and experts can have their own judgment.}
 \vspace{-2mm}
 \label{fig:annotationa_table}
\end{figure}

\par In the realm of comprehensive writing tasks, the abstract writing scoring rubric should encompass the dimensions outlined in the rubric for independent writing, while also incorporating specific requisites for abstract composition. The five dimensions inherent to independent writing encompass \textit{content integrity}, \textit{content organization}, \textit{language expression}, \textit{communicative function}, and \textit{writing conventions}~\cite{weigle2002assessing}. These dimensions collectively address various aspects of writing, ranging from coherence to linguistic proficiency. For abstract writing, the evaluation framework extends beyond the five dimensions and accentuates the importance of \textit{comprehensibility}, \textit{fluency}, and \textit{conciseness} as fundamental markers of communicative efficacy and adherence to writing conventions. In addition to these dimensions, the assessment also underscores the necessity for abstracts to exhibit alignment with the source text. The formal quality assessment of abstracts is thus underpinned by these seven key aspects~\cite{salager1994hedges}. For an elaborate exposition of the scoring criteria, kindly refer to the table (\autoref{fig:annotationa_table}) delineated in our user study.

\end{document}